\documentclass[review]{elsarticle}

\usepackage{xcolor}
\usepackage{multirow}
\usepackage{amsmath, amssymb}
\usepackage{hyperref}
\journal{Journal of Information Processing and Management}
\usepackage{array,graphicx}
\usepackage{booktabs}
\newcommand*\rot{\rotatebox{90}}

\begin{document}
	\def\hltcolor{white}

\begin{frontmatter}
	
	\title{Mining Shape of Expertise: A Novel Approach Based on Convolutional Neural Network}
	
	\author[mymainaddress]{Mahdi Dehghan}
	\ead{mah.dehghan@mail.sbu.ac.ir}
	
	\author[secondaddress]{Hossein Ali Rahmani}
	\ead{srahmani@znu.ac.ir}
	
	\author[mymainaddress]{Ahmad Ali Abin\corref{mycorrespondingauthor}}
	\ead{a\_abin@sbu.ac.ir}
	
	\author[vuaddress]{Viet-Vu Vu}
	\ead{vuvietvu@vnu.edu.vn}
		
	\cortext[mycorrespondingauthor]{Corresponding author}
	\address[mymainaddress]{Faculty of Computer Science and Engineering, Shahid Beheshti University, Tehran, Iran}
	\address[secondaddress]{Faculty of Computer Science and Engineering, University of Zanjan, Zanjan, Iran}
	\address[vuaddress]{VNU Information Technology Institute, Vietnam National University, Hanoi, Vietnam}

	\begin{abstract}
    Expert finding addresses the task of retrieving and ranking talented people \colorbox{\hltcolor}{on} the subject of user query.  It is a practical issue in the Community Question Answering networks. Recruiters looking for knowledgeable people for their job positions are the most important clients of expert finding systems. In addition to employee expertise, the cost of hiring new staff is another significant concern for organizations. An efficient solution to cope with this concern is to hire T-shaped experts that are cost-effective. In this study, we have proposed a new deep  \colorbox{\hltcolor}{model} for T-shaped experts finding based on Convolutional Neural Networks. The proposed  \colorbox{\hltcolor}{model} tries to match queries and users by extracting local and position-invariant features from their corresponding documents. In other words, it detects users' shape of expertise by learning patterns from documents of users and queries simultaneously. The proposed  \colorbox{\hltcolor}{model} contains two parallel CNN's that extract latent vectors of users and queries based on their corresponding documents and  \colorbox{\hltcolor}{join them} together in the last layer to match queries with users. Experiments on a large subset of Stack Overflow documents indicate the effectiveness of the proposed method against baselines in terms of NDCG, MRR, and ERR evaluation metrics.
    \end{abstract}
	
	
\end{frontmatter}

%\linenumbers

\section{Introduction}
\label{introduction}
Community Question Answering (CQA) networks such as Stack Overflow\footnote{stackoverflow.com}, Quora\footnote{quora.com}, and Yahoo! Answers\footnote{answer.yahoo.com} have gained \colorbox{\hltcolor}{much} popularity and interests among people due to providing a valuable interface to share or exchange knowledge and information. This growing popularity makes people contribute more in such networks to discuss their questions or respond to other users' questions. In addition, people can vote up or down questions or answers to specify their importance and quality in such networks. Also, there is an ability for the questioner to determine the best answer of its question in some CQAs. The huge source of users' generated information in CQA networks, have attracted attention of many researchers to define challenging and practical problems such as expert finding \cite{wang2018survey}, question routing \cite{mumtaz2019expert2vec,zhou2009routing}, question classification \cite{momtazi2018unsupervised,mohasseb2018question}. Among all problems, expert finding with the aim of detecting and ranking talented people in the subject of user's query is a well-studied problem in the field of Information Retrieval (IR). Expert finding as a well-studied problem has been investigated in other domains such as bibliographic networks \cite{hashemi2013expertise}, microblogs and social networks \cite{neshati2013joint}. Finding knowledgeable people has \colorbox{\hltcolor}{a} large number of real-world applications. Detecting experts and offering them to recruiters who seek talented people in a specific skill area is one of the most important applications of expert finding. 

Stack Overflow as one of the most popular and active CQAs, provides its users a good platform to share knowledge and seek information. Figure \ref{fig:StackOverflow} demonstrates a typical page of Stack Overflow in which important informative parts are highlighted. %In Stack Overflow, users can post questions or answers, vote up or down them and leave comments. The questioner can mark the best answer of its question as accepted answer.
In Stack Overflow, the questioner \colorbox{\hltcolor}{has} to determine one or more tags to its question. Tags \colorbox{\hltcolor}{indicate} the skills needed to respond the question and are used as \colorbox{\hltcolor}{query terms} in expert finding problem. As mentioned, linking talented people to companies intending to hire new staff is the most significant application of expert finding in CQA networks. Human resources are a valuable property to companies since they possess a range of skills which can benefit the companies. Recruiters prefer to hire new staff with the lowest cost. Employees with high range of expertise areas (i.e. C-shaped expert) demand high salary. Therefore, mining shape of expertise helps companies to find appropriate person with reasonable cost for their job positions. Gharebagh et al.~\cite{gharebagh2018t} have categorized users of Stack Overflow into three: T-shaped, C-shaped and non-expert groups. T-shaped users have deep knowledge or skills in one skill area and a broad base of general supporting knowledge or skills. C-shaped users have deep knowledge or skills in several skill areas. \colorbox{\hltcolor}{Non-expert} users  \colorbox{\hltcolor}{don't have} deep knowledge or skills in any skill area.  

\def\sc{.3}
\begin{figure}
	\begin{center}
		\includegraphics [scale=\sc] {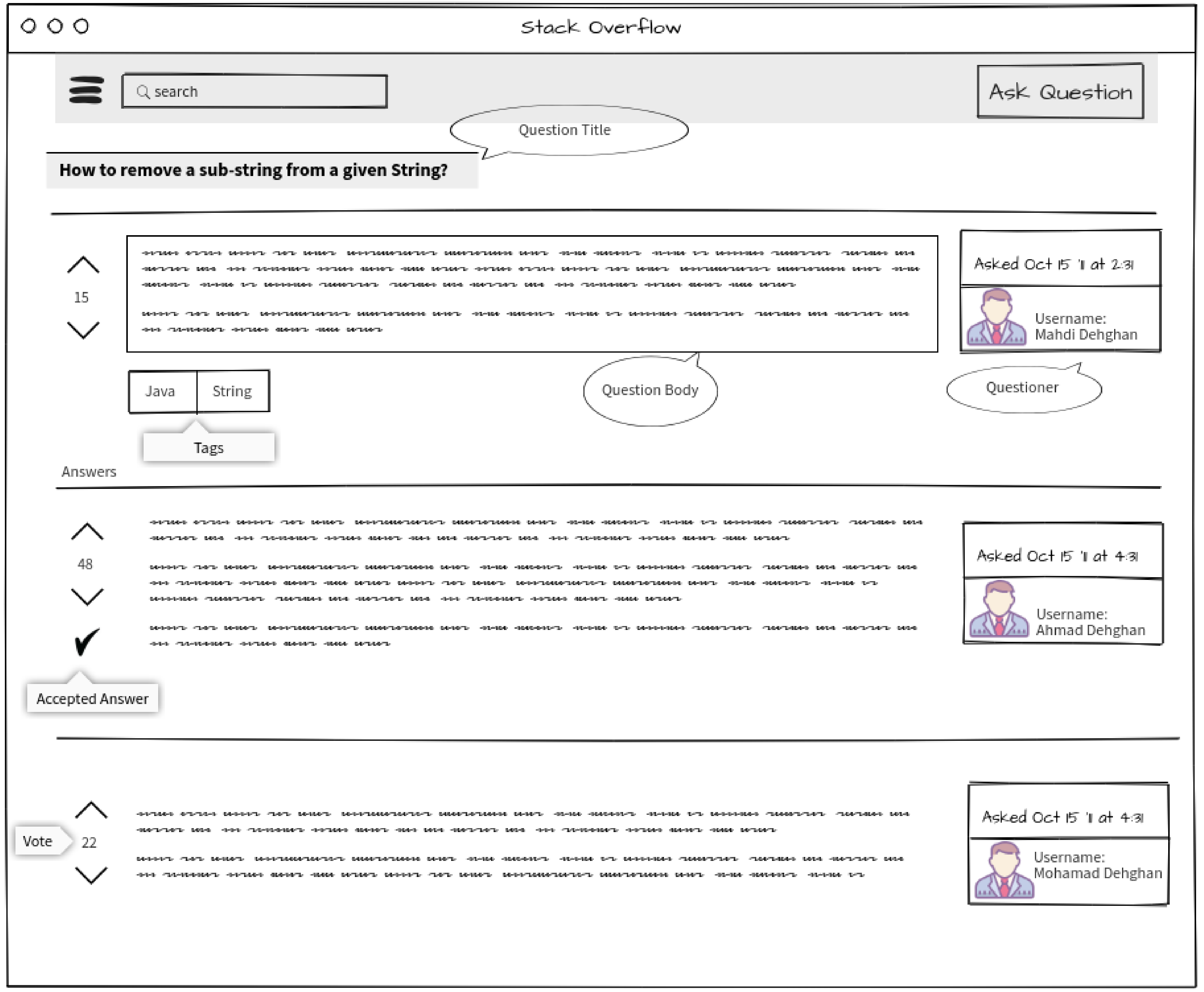}
	\end{center}
	\caption{A typical page of Stack Overflow.}
	\label{fig:StackOverflow}
\end{figure}

Nowadays, agile methodology has attracted attention of many companies for software development. Mining shape of expertise helps companies to form an agile software  development team with reasonable cost. Among all strategies to form an agile team, the best one forms a team with the lowest cost and the highest performance simultaneously. Consider a team in which all members are \colorbox{\hltcolor}{C-shaped experts or full-stack developers} in different aspects of the project. Although team formation by this strategy brings about a team with high performance, it imposes \colorbox{\hltcolor}{much cost}to companies because C-shaped experts \colorbox{\hltcolor}{always ask for higher salaries}. The best way of forming an agile team is to hire T-shaped experts based on aspects of the project. In this way, each team member covers a particular  \colorbox{\hltcolor}{aspect} of the project and have general knowledge of other aspects of the project. This will preserve the principles of modern software engineering up to date, and all team members will be involved in all aspects of the project if necessary \cite{gharebagh2018t}. Therefore, T-shaped expert finding is a practical and industry-motivated problem in the field of IR \cite{gharebagh2018t, dehghan2019temporal}.

In this paper, we have proposed a novel deep neural network based architecture to model both users expertises and skill areas jointly from their corresponding documents. In other words, we have tried to find a matching between skill areas and  users semantically. The proposed method consists of two  parallel Convolutional Neural Networks that are joined in the last layers. One CNN extracts user expertise vector form his documents, and the other one extracts vector for skill area from existing documents in  \colorbox{\hltcolor}{a} specified skill area. These two parallel networks are joined in the last layer of the proposed architecture. Finally, outputs of the joint layer are connected to an output neuron for expertise shape detection.

The rest of this paper is organized as follows. Research objective and contribution of this work are highlighted in Section \ref{contribution_of_this_work}. Related works are discussed in Section \ref{related_work}. In Section \ref{proposed_method}, we present the proposed method for T-shaped expert finding. Details for the dataset, baseline models, evaluation measures, parameters setting, and implementation are given in Section \ref{experimental_stup}. Finally, the paper concludes with conclusions and future directions in Section \ref{conclusion}.

\section{Contribution of this work}\label{contribution_of_this_work}
T-shaped expert finding aims to detect and rank T-shaped experts in the subject of user's query. This practical problem has great application in forming agile software development teams. In this study, we have proposed a new deep architecture for T-shaped experts finding based on Convolutional Neural Networks. The proposed architecture tries to match queries and users by extracting local and position-invariant features from their corresponding documents. In other words, it detects users' shape of expertise by learning patterns from documents related to users and queries simultaneously. As a consequence, it can relieve the vocabulary gap problem to some extent. Unlike  the method proposed in \cite{gharebagh2018t}, our proposed method can predict both C- and T-shaped ranking of experts based on user's query. Also,  we predict \colorbox{\hltcolor}{the shape} of expertise for each user by extracting text features from his documents. \colorbox{\hltcolor}{Moreover}, \colorbox{\hltcolor}{this study addresses the main issue} of the method proposed in \cite{dehghan2019temporal} that is not\colorbox{\hltcolor}{ feasible with large datasets}. In fact, it is a common problem in many learning based expert profiling methods because all queries have to be determined in advance. To sum up, we have proposed a CNN-based model which detects different shapes of expertise by learning \colorbox{\hltcolor}{the latent} vectors of users and queries \colorbox{\hltcolor}{in} Stack Overflow.

\section{Related Work}\label{related_work}
In this section, we first \colorbox{\hltcolor}{define the expert finding problem} and  \colorbox{\hltcolor}{outline} some of \colorbox{\hltcolor}{its applications}. Then, we review some prior studies on community question answering. Finally, we \colorbox{\hltcolor}{survey} some important studies in mining shape of expertise.

\subsection{Expert Finding}
Expert finding that addressed the task of detecting and ranking right people in the subject of user's query, has become an important issue in the field of IR.  
Over the past years, many research studies have conducted to solve this problem in different domains, each of which tried to improve the quality of ranking by dealing with a specific challenge \cite{balog2012expertise,wang2018survey}. In the past, expert finding methods were categorized into two graph-based \cite{erten2004exploring,mutschke2003mining,white2003algorithms,zhang2007expertise} and topic modeling- based \cite{momtazi2013topic,zhou2012topic,li2015hybrid} groups. Over time, researchers have proposed probabilistic methods for expert finding \cite{balog2009language}. Finding talented people has found many applications in different environments such as CQA networks \cite{neshati2017dynamicity,zhou2016learning,zhao2016expert}, bibliographic networks \cite{review2_2,review2_1}, academic \cite{alarfaj2012finding,deng2008formal}, organizations \cite{karimzadehgan2009enhancing}. Balog et al. \cite{balog2009language} have proposed two outstanding language models for expert finding. Their language models were named candidate-based and document-based model. In candidate-based model, for each candidate expert a textual representation is \colorbox{\hltcolor}{built} upon his associated documents. In fact, candidate-based model represents a candiate expert as a language model, then the probability of generating query topic given this candidate (i.e. its language model) is considered as candidate's level of expertise. In the document-based model, first, the probability of generating a user's query (referred to as document value) is calculated for any document in the collection. Then, each candidate, according to his contribution in the document, obtains the score from that document. In the end, the candidates are sorted by their total scores.

\subsection{Expert Finding in CQA Networks}
Over the recent years, many research studies have been conducted to solve different challenging problems on CQA networks \cite{wang2018survey,mumtaz2019expert2vec,mohasseb2018question}. In CQA networks, documents are posted by registered users in the forms of questions and answers. In such networks, a questioner posts a question, one ore more answerers reply the question, and finally the questioner marks an answer as accepted answer. Also, the questioner \colorbox{\hltcolor}{has} to tag his question \colorbox{\hltcolor}{with} some keywords indicating the skills needed to respond \colorbox{\hltcolor}{to} a question. CQA networks provide valuable source of information for expert finding researchers. Expert finding in CQA networks is challenging task due to query term mismatch or vocabulary gap problem \cite{Abin2019TJEE, dehghan2019translations, dargahi2017skill, li2014semantic}.   

Dargahi et al.~\cite{dargahi2017skill} have conducted a research study to investigate the efficiency of translation models in relieving the query term mismatch problem in CQA networks. Their translation models were named mutual information-based (referred to as MI-based) and word embedding-based (referred to as WE-based) approach. MI \colorbox{\hltcolor}{measure}  \colorbox{\hltcolor}{shows} the relevancy of words \colorbox{\hltcolor}{according to their} \colorbox{\hltcolor}{co-occurrence} in a document. In the MI-based approach, they have used Normalized MI as translation probability. In their WE-based approach, they have proposed a neural network structure to model the relation between topics and queries to \colorbox{\hltcolor}{decrease} the gap between query and words. Eventually, they have utilized query translations in a binary scoring based approach to find experts. Authors in \cite{nobari2020quality} have utilized other features to improve the results. Accordingly, they have introduced the concept of vote share to determine the quality of the posts. Eventually, the posts with higher vote share would have more impact on the translation model.

Although proposed translation models in \cite{dargahi2017skill,nobari2020quality} can bring about an improvement in the quality of expert finding, they did not consider the diversification concept that can cause to cover more query topics and improve the quality of final expert ranking. Dehghan et al.~\cite{dehghan2019translations} have proposed a translation model taking the diversification concept into account to cope with the vocabulary gap problem in CQA networks. Their translation model is based on data clustering in two query and co-occurrence spaces. Their proposed translation model clusters all important words extracted by topic modeling in \colorbox{\hltcolor}{the} query space at the first step and then clusters each query space cluster in the co-occurrence space again to remove semantically similar words. Then, they translate each query into a set of relevant and diverse words through a probabilistic framework. Finally, they utilize the query translations in a voting-based approach to identify experts. 

Topic modeling is one of the semantic matching \colorbox{\hltcolor}{approaches} utilized to deal with \colorbox{\hltcolor}{the } query term mismatch problem. Momtazi et al.~\cite{momtazi2013topic} have proposed a topic modeling based approach to retrieve and rank experts on the Text Retrieval Conference (TREC) Enterprise track for 2005 and 2006. They have extracted main topics of queries by using Latent Dirichlet Allocation (LDA) \cite{blei2003latent} in the first step. Then, they have utilized the topics that are extracted from their document collection as a bridge to connect candidate experts to user's query. Van Gysel et al.~\cite{van2016unsupervised} have introduced an unsupervised neural network based approach to solve \colorbox{\hltcolor}{an} expert finding task. Their model learns latent vectors of candidate experts and queries  based on the occurrence of query words in the documents posted by candidate experts. 

Temporal property of expertise is another important feature that has been investigated by researchers in CQA networks. Neshati et al.~\cite{neshati2017dynamicity} have introduced future expert finding problem in CQA networks, which is the task of predicting talented users in the future according to their expertise evidence \colorbox{\hltcolor}{at} the current time. They have proposed two learning algorithms in which they have investigated the impact of four different feature  groups, including topic similarity, emerging topics, user behaviour, and \colorbox{\hltcolor}{topic transition in} the future expert finding task. Patil et al.~\cite{patil2016detecting} have proposed a method to detect experts in Quora. They have analysed the behaviour of both expert and non-expert users in Quora according to four feature \colorbox{\hltcolor}{groups, namely:} user activity, quality of answers, linguistic and temporal. Then, they have developed statistical classification models in order to detect expert people by choosing high-resolution features.

\subsection{Mining Shape of Expertise}
Up to now, we have discussed some previous works in expert finding that only concentrated on detecting expert people. Authors in \cite{dehghan2019temporal,gharebagh2018t} have proposed methods to detect \colorbox{\hltcolor}{the} shape of expertise in CQAs networks. Gharebagh  et  al.~\cite{gharebagh2018t} have built a dataset of Stack Overflow users in which all users are categorized into three T-shaped, C-shaped and non-expert groups according to their knowledge and skills in \colorbox{\hltcolor}{the} extracted skill areas. Then, they have proposed a probabilistic model to find T-shaped experts. Dehghan et al.~\cite{dehghan2019temporal} have proposed an outstanding temporal expert profiling model on Stack Overflow. They have conducted their study to prove \colorbox{\hltcolor}{the} effect of time in expertise. In fact, they have introduced an LSTM neural network based approach to model the relation between temporal expertise trees and shape of expertise.

\section{Proposed Method}
\label{proposed_method}
In this section, the proposed method is discussed in more detail. Figure \ref{fig:model} shows the idea behind the proposed method. As this figure shows, the proposed method, jointly models both user expertise and query properties by using \colorbox{\hltcolor}{the} answers in Stack Overflow. It learns latent vectors of users and queries from their corresponding answers such that the learned vectors can predict the user's shape of expertise. This process is done by a CNN-based model consisting of two parallel convolution neural networks (CNN) joined with together in the last layer. The joint model is then trained in a supervised manner to predict the user's level of expertise with minimum prediction error. In fact, the proposed method outputs a ranking of users for a given query based on their shape of expertise. In this work, we intend to create a ranking of T-shaped experts. There, we adjust the output neuron of the model to indicate the level of being a T-shaped expert. In other words, we intend to solve the problem of T-shaped expert finding. However, our proposed method can also propose a ranking of C-shaped experts by a slight modification in \colorbox{\hltcolor}{the} training phase (Section \ref{constructing_golden_set}). In the following, we first describe notations used in this paper. Then, the architecture of the proposed model is explained. Finally, we illustrate how to train the proposed model.

\def\sc{.8}
\begin{figure}
	\begin{center}
		\includegraphics [scale=\sc] {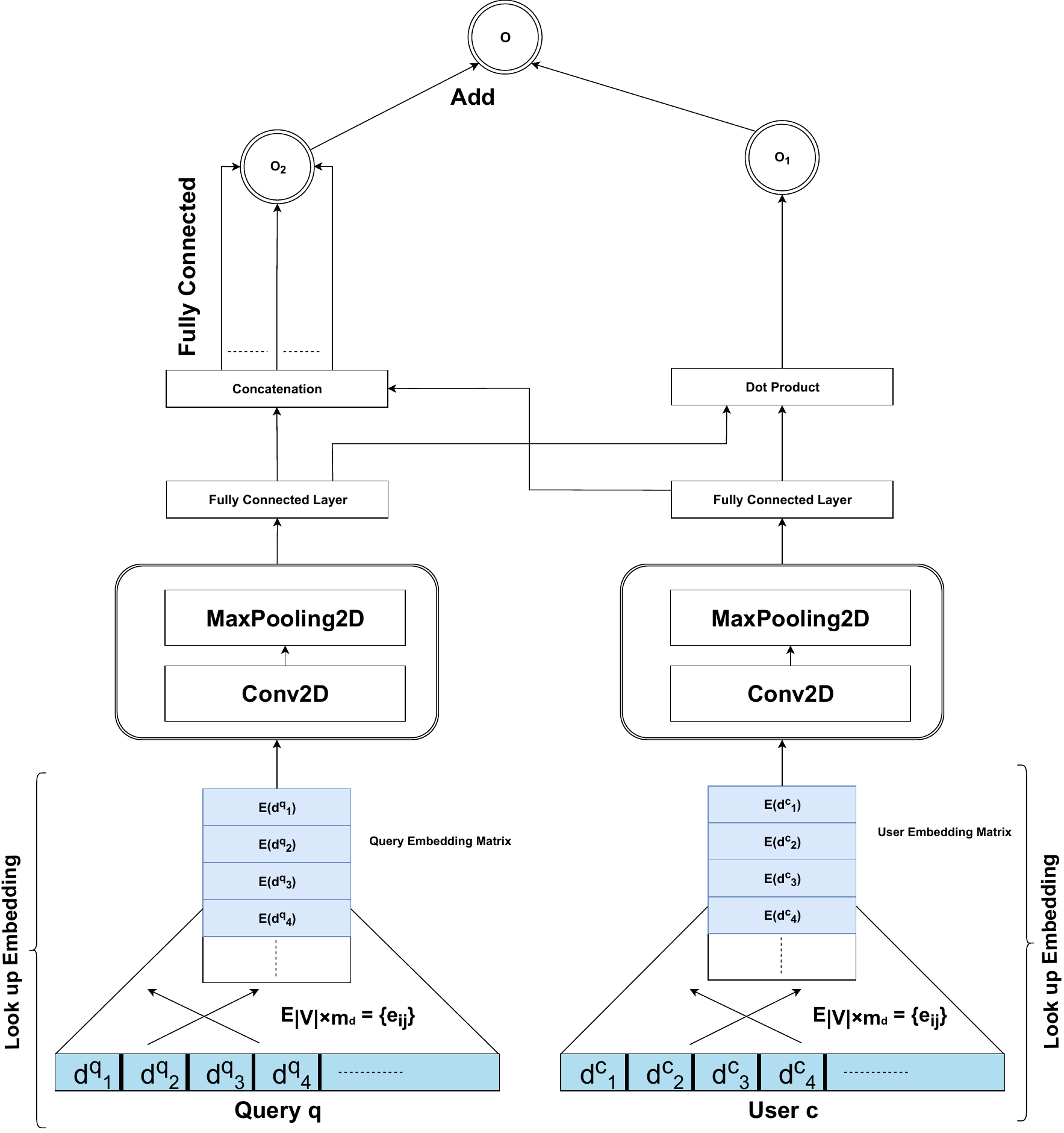}
	\end{center}
	\caption{The proposed CNN-based approach for detecting shape of expertise.}
	\label{fig:model}
\end{figure}

\subsection{Preliminaries}
\label{preliminaries}

In the setting of this work, we have a collection of document (i.e. answers in Stack Overflow) $ D = \{d_1^{\bullet}, d_2^{\bullet}, ..., d_x^{\bullet}\} $, set of candidate experts $ C = \{c_1, c_2, ..., c_y\} $, and set of queries $ Q = \{q_1, q_2, ..., q_z\} $. We let $ D^{c_i} $ and $ D^{q_j} $  denote set of documents posted by candidate expert $ c_i \in C $, and set of documents related to query $ q_j $, respectively. Table \ref{table:notations} indicates the summary of notations used in this work. In the following, we explain the proposed method in more detail.

\begin{table}
\centering
\caption{Summary of notations.}
	\begin{tabular}{ll}
		\hline
		\textbf{Notation} & \textbf{Notation description}\\
		\hline
		$ {q_{\bullet}} $ & Query\\		
		\hline
		$ \boldsymbol{c_{\bullet}} $ & Candidate expert\\
		\hline
		$ \boldsymbol{D} $ & Document collection (i.e. Stack Overflow answers)\\
		\hline
		$ \boldsymbol{C} $ & Set of all candidate experts\\
		\hline
		$ \boldsymbol{Q} $ & Set of all queries\\
		\hline
		$ \boldsymbol{D^{c_i}} $ & Set of documents posted by candidate expert $ c_i $\\
		\hline
		$ \boldsymbol{D^{q_j}} $ & Set of documents related to query $ q_j $\\
		\hline
		$ \boldsymbol{m_d} $ & Size of document embedding vector\\
		\hline
		$ \boldsymbol{\chi_c} $ & Convolutional neural network layer in the proposed model for candidate experts\\
		\hline
		$ \boldsymbol{\chi_q} $ & Convolutional neural network layer in the proposed model for queries\\
		\hline
		$ \boldsymbol{m_c} $ & Size of candidate latent vector (i.e. the number of output neurons of $ \chi_c $)\\
		\hline
		$ \boldsymbol{m_q} $ & Size of query latent vector (i.e. the number of output neurons of $ \chi_q $)\\
		\hline
		$ \boldsymbol{E_{\bullet}} $ & Document embedding matrix\\
		\hline
		$ \boldsymbol{n} $ & Maximum number of documents that have to select from $ D^{c_i} $ and $ D^{q_j} $\\
		\hline
		$ \boldsymbol{k} $ & Number of convolution filters in $ \chi_c $ and $ \chi_q $\\
		\hline
		$ \boldsymbol{f} $ & Number of rows in convolution filters\\
		\hline
		$ \boldsymbol{p} $ & Size of feature map in max pooling kernel\\
		\hline
		$ \boldsymbol{L_{\bullet}} $ & Latent vector\\
		\hline
	\end{tabular}
	\label{table:notations}
\end{table}

\subsection{Problem Specification}\label{problem_specification}
Given query $ q_j \in Q $, T-shaped expert finding task is to return a ranked list of candidate experts $C_p = \{c_i \in C | c_i \text{ is a T-shaped expert in the query } q\}$, $|C_p| = R$,in the given query. To do so, we propose a deep architecture based on two parallel CNNs that learns latent vectors of users and queries in a supervised manner. For each query $q$, we feed all $(D^{q}, D^{c_i})$ pairs as the inputs to the proposed model. The proposed model assigns a value to each pair (i.e. $o_3$) indicating the level of being T-shaped expert. Therefor, $C_p$ is generated by sorting all users based on the predicted value by model.

\subsection{Model Architecture}
\label{model_architecture}

Figure \ref{fig:model} demonstrates the architecture of the proposed model for predicting level of being a T-shaped expert. As mentioned before, the model is composed of two parallel convolutional neural networks joined in the last layers, $ \chi_c $ and $ \chi_q $ for candidate experts and queries, receptively. In the model, the documents posted by candidate expert $ c_i $, $ D^{c_i} $, and the documents related to query $ q_j $, $ D^{q_j} $, are fed to $ \chi_c $ and $ \chi_q $ respectively as inputs, and level of being T-shaped expert is produced as the output. In the first layer, referred to as look-up embedding layer, $ D^{c_i} $ and $ D^{q_j} $ are represented as matrices of document embedding to preserve the semantic information in the documents. The next layers of the model are the common CNN layers consisting of convolution, pooling, and fully connected layers utilized to extract high-resolution feature vector for candidates and queries. At the top of the model, the output of $ \chi_c $ and $ \chi_q $ are joined together to connect latent vectors of candidates and queries. Eventually, the output layer consists of a single neuron that minimizes the level of being T-shaped expert based on latent vectors $ \chi_c $ and $ \chi_q $. In the following, we explain how to extract $ \chi_c $. Extraction of  $ \chi_q $ is same as $ \chi_c $.

\subsubsection{Look up Embedding Layers}
\label{look_up_embedding_layers}

A document embedding $ \varphi : D \rightarrow {\rm I\!R}^{m_{d}}$ is a parametrized mapping of documents into $ m $-dimensional vectors. Recently, this approach has found many applications in a large number of text mining tasks \cite{markov2016author,maslova2017neural}. We utilize LDA \cite{blei2003latent} to extract semantic vectors of documents. In the look up embedding layer, each candidate $ c_i \in C $ has a set of posted documents $ D^{c_i} $ that is represented as a matrix of document embeddings $ E_{c_i} \equiv [e_{xy}]_{n \times m_d} $. Since $ E_{\bullet} $ should be square matrix, we just select $ n $ documents for each candidate expert. For candidates having less than $ n $ documents, we pad the end of matrix by zeroes. \colorbox{\hltcolor}{The following equation} illustrates how the look up embedding layer works:

\begin{equation}\label{eq:look_up_embeeding_layer}
E_{c_i}^{1 \le j \le n} = \left\{\begin{array}{ll} \varphi(d_{j}^{c_i} \in D^{c_i}) & if\ j \leq \mid D^{c_i} \mid \\ {[0, 0, 0, ..., 0]}_{1 \times m_d} & otherwise \end{array}\right.
\end{equation}

To sum up, look up embedding layer receives set of documents posted by candidate experts and passes a matrix of document embedding to the next layer (i.e. CNN layer).

\subsubsection{CNN Layers}
\label{cnn_layers}

Next layers consist of convolution, max pooling, and fully connected layers. A convolution layer consists of $ k $ filters, $ F_j \in {\rm I\!R}^{f \times m_d} \ j=1,...,k$, that generates new features by doing convolution operator on matrix embedding $ E_{\bullet} $. Each vector generated by a filter is passed to the max pooling layer. \colorbox{\hltcolor}{The} max pooling layer selects the maximum value as the most important feature of each feature map with a size of $ p $. In fact, it performs a down sampling operation along the spatial dimensions of its inputs. After max pooling layer, there is a fully connected layer that receives the results of max pooling layer. The fully connected layer produces a vector with size of $ m_{c} $ indicating latent vector for a candidate expert $ c_i $ (i.e. $ L_{c_i} $). As mentioned earlier, $ \chi_q $ and $ \chi_c $ have the same structure. So, the latent vector of query $ q_j $ with size of $ m_{q} $ (i.e. $ L_{q_j} $) are produced as the outputs of $ \chi_q $. 

\subsubsection{Integration Layers}
\label{shared_layers}

Having latent vectors of candidate expert $ c_i $, $ L_{c_i} $, and query $ q_j $, $ L_{q_j} $, we intend to predict the level that the candidate expert $ c_i $ is a T-shaped expert in $ q_j $. To this end, we have utilized these vectors in two different ways: 1) their dot product is connected to a single neuron $ o_1 $ and, 2) their concatenation is connected to a single neuron $ o_2 $. Finally, the output neuron $ o_3 $ returns the level that  candidate $ c_i $ is T-shaped expert in query $ q_j $ by using $ o_1 $ and $ o_2 $.

\subsection{Constructing Golden Set}
\label{constructing_golden_set}

In this section, we explain how to create training and testing datasets in order to train and evaluate the network. As mentioned, the inputs to the network are set of documents posted by a candidate expert $ c_i $ and set of documents related to query $ q_j $, and the desired output is the scalar $ o_3 $ indicating the level that the candidate expert $ c_i $ is a T-shaped expert in query $ q_j $. To train the network, we have to know the true value of $ o_3 $ for each input pair $ (D^{c_i}, D^{q_j}) $. Knowing the shape of each candidate expert $ c_i $ in our dataset, we determine \colorbox{\hltcolor}{the} true value of $ o_3 $ in the following way: 
\begin{enumerate}
	\item If the candidate expert $ c_i $ is a T-shaped expert in query $ q_j $, $ o_3 $ will be assigned to $ 1 $.
	\item If the candidate expert $ c_i $ is a C-shaped expert in query $ q_j $, $ o_3 $ will be assigned to $ 0 $.
	\item If the candidate expert $ c_i $ is a non-expert in query $ q_j $, $ o_3 $ will be assigned to $ -1 $.
\end{enumerate}
By these assignments, the output neuron $ o_3 $  indicates the level that a candidate expert $ c_i $ is a T-shaped expert in query $ q_j $. If we intend to predict the level of being C-shaped experts, we have to assign $ 0 $ to the output neuron $ o_3 $. After constructing network inputs and outputs by this way, $ 60\% $ of the available data for T-shaped, C-shaped and non-expert users are allocated for training. The remaining $ 40\% $ data are equally partitioned and referred to as validation and test data sets.

\subsection{Ranking Candidate Experts}
\colorbox{\hltcolor}{The ranking of candidate experts} is the final step in the proposed method. Since $ o_3 $ indicates the level of being T-shaped expert, to find a ranking of T-shaped experts in query $ q_j $, \colorbox{\hltcolor}{all candidate experts will be ordered} by their corresponding $ o_3 $ and return the ranked list of T-shaped experts in query $ q_j $.

\section{Experimental Setup}
\label{experimental_stup}

In this section, we compare the proposed method with the baselines by using three well-known evaluation metrics in the IR community. In the following, we present details of evaluation metrics, parameters settings, and implementations.

\subsection{Dataset} \label{dataset}
We use a real-world CQA dataset from Stack Overflow provided by \cite{gharebagh2018t}. This dataset crawled from Stack Overflow contains all \colorbox{\hltcolor}{posts} from August $ 2008 $ to March $ 2015 $ and consists information of about $ 4 $ million users and $ 24,120,523 $ posts \cite{gharebagh2018t}. The dataset consisting of three separate data collections has been generated from all posts tagged \colorbox{\hltcolor}{with} $ Java $, $ Android $, and $ C\# $. Each domain includes a number of skill areas or queries each of which consists of a number of tags. We categorized each user in one of T-shaped, C-shaped, and non-expert group based on his knowledge level. In the following, we explain the way of extracting queries for a domain and labeling users as T-shaped, C-shaped, and non-expert.

\subsection{Extraction of Skill Areas}\label{skill_area}
As mentioned, a \colorbox{\hltcolor}{query consists of a set of related tags}. Gharebagh et al. \cite{gharebagh2018t} did three steps to extract queries for a particular domain. At first, they have extracted $ 200 $ \colorbox{\hltcolor}{most frequent} tags of the domain. In the second step, they used an agglomerative clustering \colorbox{\hltcolor}{algorithm} to cluster the extracted tags. Finally, a number of computer industry artisans revised and refined \colorbox{\hltcolor}{the} clusters based on the interest and need for recruitment to label them as a query. Table \ref{tab:skill_arae} shows some queries in Java, Android and C\# domains.
\begin{table}
	\caption{Some skill areas or queries in Java, Android, and C\# domains \cite{dehghan2019temporal}.}
	\centering
	\small
	\begin{tabular}{lll}
		\noalign{\hrule height 1pt}
		\textbf{Domain} & \textbf{Query} & \textbf{Tags}\\
		\noalign{\hrule height 1pt}
		Java & Spring & spring, spring-mvc, spring-security \\
		\hline	
		Android & Push-notifications & android-gcm, push-notifications, android-notifications \\
		\hline
		C\# & Unit-testing & moq, unit, unit-testing\\
		\noalign{\hrule height 1pt}
	\end{tabular}
	\label{tab:skill_arae}
\end{table}

\subsection{\colorbox{\hltcolor}{Classification of users}}\label{categorization_of_users}
In this section, we introduce how to classify users into one of T-shaped, C-shaped, or non-expert group by using two measures named precision and recall. 
\begin{enumerate}
	\item $ Precision(q_i, c_j) $: The ratio of the number of accepted answers posted by user $ c_j $ in query $ q_i $ to the total number of his answers in query $ q_i $.
	\item $ Recall(q_i, c_j) $: The ratio of the number of accepted answers posted by user $ c_j $ in query $ q_i $ to the total number of accepted answers in query $ q_i $.
\end{enumerate}

To classify users, firstly, we rank them based on \colorbox{\hltcolor}{the} harmonic mean of precision and recall. The top $ 5\% $ of the ranked users are labelled as advanced knowledge level users. The next $ 20\% $ of them are labelled as intermediate knowledge level users. The rest of the list are considered to be  beginner knowledge level users. Finally, we classify each user in one of T-shaped, C-shaped or non-expert groups based on his knowledge level by using the following rules:
\begin{enumerate}
	\item Non-expert user has no advanced knowledge level in any skill area.
	\item T-shaped expert user has advanced knowledge level in only one skill area and intermediate knowledge level in at least one skill area.
	\item C-shaped expert user has advanced knowledge level in more than one skill area.
\end{enumerate}

Table \ref{table:dataset} indicates the statistics of the dataset. Also, Figure \ref{fig:share_expertise} plots share of different expertise shapes in Java , Android and C\# domains.

\begin{table}
	\caption{General statistics of dataset.}
	\centering
	\small
	\begin{tabular}{lcccccc}
		\noalign{\hrule height 1pt}
		\textbf{Domain} & \textbf{\#Questions} & \textbf{\#Answers} & \textbf{\#Queries} & \begin{tabular}[x]{@{}c@{}} \textbf{\#All} \\ \textbf{users} \end{tabular} & \begin{tabular}[x]{@{}c@{}} \textbf{\#T-shaped} \\ \textbf{users} \end{tabular} & \begin{tabular}[x]{@{}c@{}} \textbf{\#C-shaped} \\ \textbf{users} \end{tabular}\\
		\noalign{\hrule height 1pt}
		$ Java $ & $ 810,071 $ & $ 1,510,812 $ & $ 26 $ & $ 83,557  $ & $ 2465 $ & $ 1673 $\\
		
		$ Android $ & $ 638,756 $ & $ 917,924 $ & $ 18 $ & $ 57,889  $ & $ 1830 $ & $ 1074 $\\
		
		$ C\# $ & $ 763,717 $ & $ 1,453,649 $ & $ 23 $ & $ 84,095  $ & $ 2707 $ & $ 1783 $\\
		\noalign{\hrule height 1pt}
	\end{tabular}
	\label{table:dataset}
\end{table}

\def\sc{0.16}
\begin{figure}
	\centering
	\begin{tabular}{ccc}
		\includegraphics [scale=\sc] {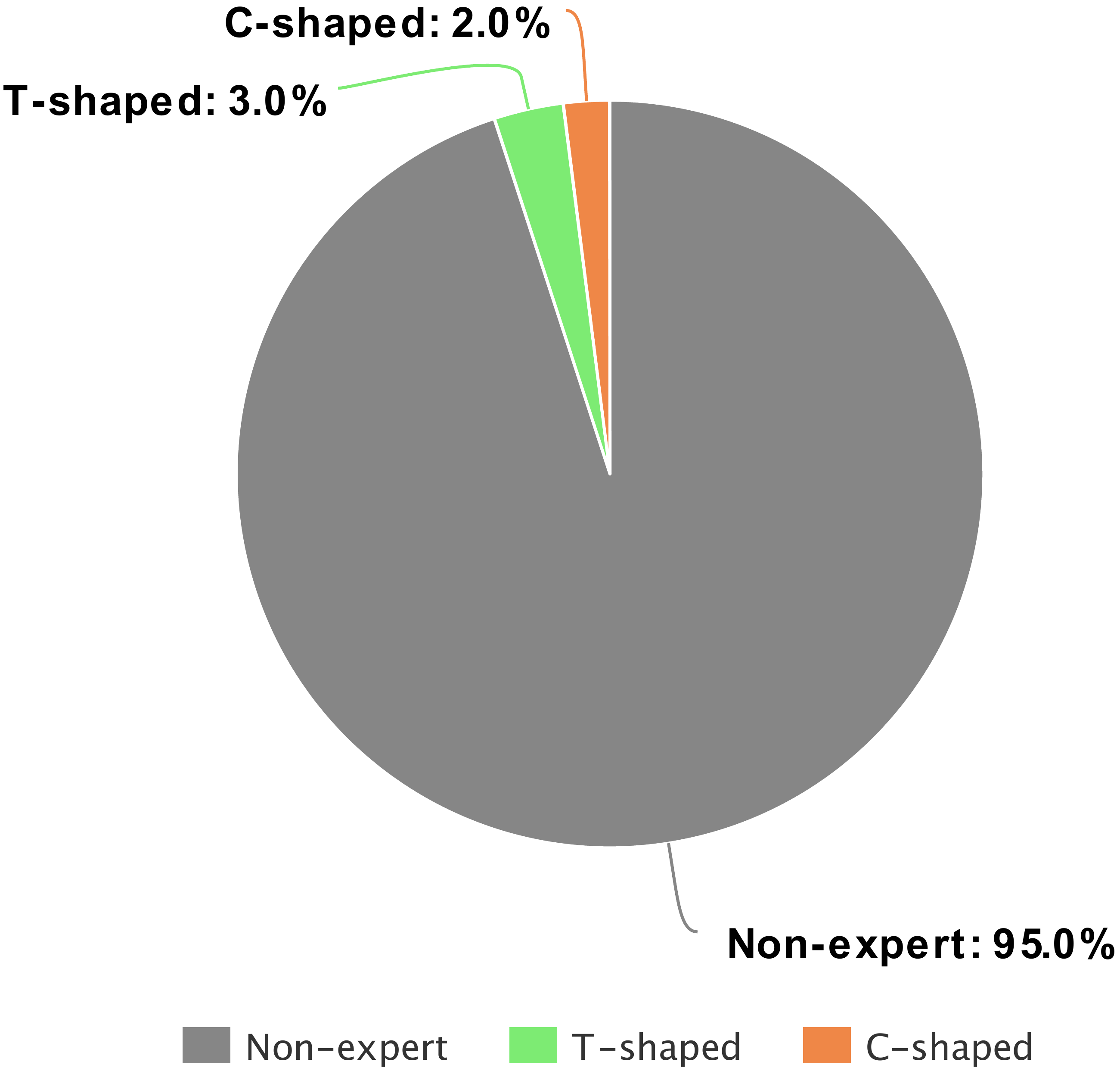} & \includegraphics [scale=\sc] {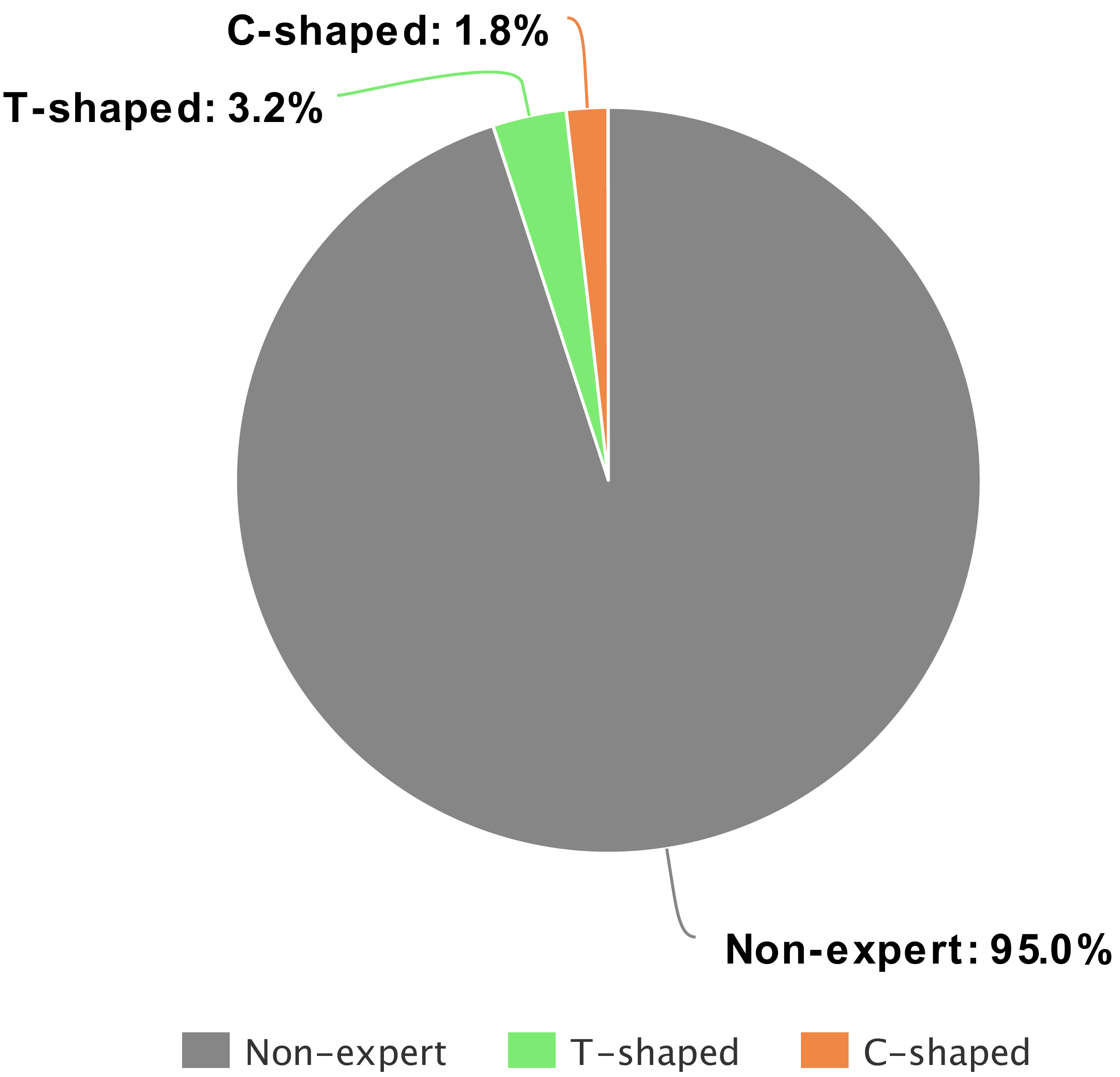} & \includegraphics [scale=\sc] {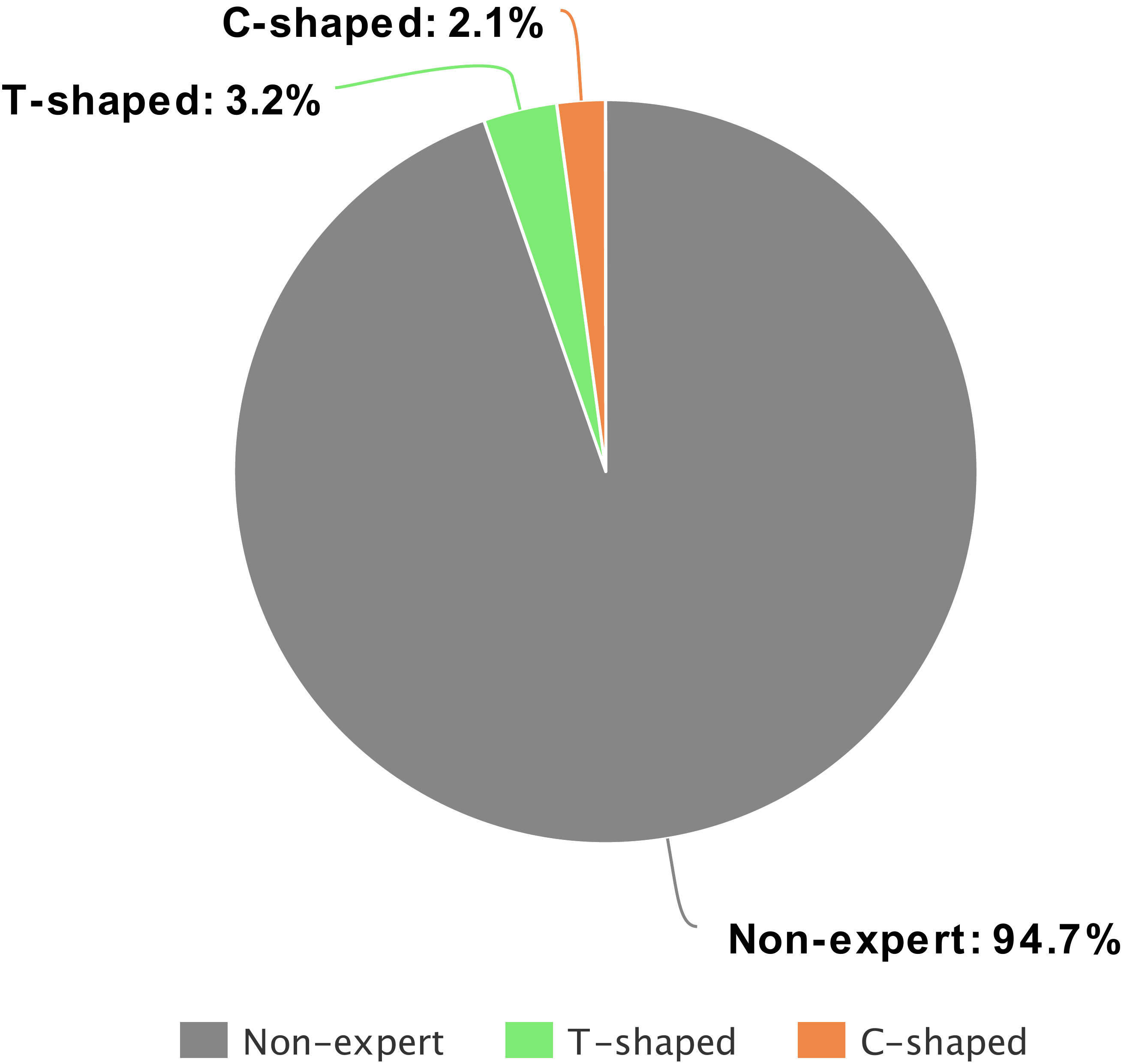}\\
		(a) Java  & (b) Android  & (c) C\# \\
	\end{tabular}
	\caption{Share of different shapes of expertise in Java, Android and C\# domains.}
	\label{fig:share_expertise}
\end{figure}

\subsection{Baseline Models}
\label{Baseleline_models}
In this section, we introduce five well-known methods in expert finding literature as our baselines. The first one is a \underline{d}ocument-\underline{b}ased \underline{a}pproach \cite{balog2012expertise} referred to as DBA. The second and third baselines are \underline{e}ntropy-\underline{b}ased \underline{a}pproach and e\underline{x}tended \underline{e}ntropy-\underline{b}ased \underline{a}pproach referred to as EBA and XEBA, respectively \cite{gharebagh2018t}. Another baseline, proposed in \cite{dehghan2019temporal}, is a temporal expert profiling method utilizing LSTM neural network to find shape of expertise. Two different traversal approach, DFS and BFS, have been utilized to train LSTM neural network in \cite{dehghan2019temporal}. Therefore, our forth and fifth baselines are named as LSTM-BFS and LSTM-DFS approaches in the rest of this paper.
\subsection{Evaluation Metrics}
\label{evaluation_measures}
In this section, we introduce three well-known evaluation metrics utilized to compare the proposed method against baselines. The first one is Normalized Discounted Cumulative Gain (NDCG). To evaluate the quality of ranking by using NDCG, we have to assign a relevancy score to each retrieved item (i.e. users). Here,  \colorbox{\hltcolor}{we assume $ 2 $, $ 1 $ and $ 0 $ as relevancy scores for T-shaped}, C-shaped, and non-expert users, respectively. The second  \colorbox{\hltcolor}{measure} is Mean Reciprocal Rank (MRR). MRR measure indicates the first occurrence of the true retrieved item in the ranking on average. The third one is Expected Reciprocal Rank (ERR) \cite{chapelle2009expected} inspired by cascade models. ERR is the expected reciprocal amount of time that any user spends on finding a relevant result \cite{dehghan2019temporal}.

\subsection{Parameters Setting and Implementation Details}
\label{parameter_setting_and_implementation_details}
We have used Keras\footnote{https://keras.io} to implement the proposed neural network model. We have used Gensim Doc2vec package\footnote{https://radimrehurek.com/gensim/} that contains an implementation of doc2vec model proposed in \cite{le2014distributed}, to extract semantic feature vector with size of $ 100 $ dimension (i.e. $ m_d = 100 $) of our documents. Also, to investigate the effect of document embedding method on the efficiency of the proposed method, we have used LDA topic modelling approach \cite{blei2003latent} to extract semantic feature vector of documents. In this work, the well-known MALLET\footnote{http://mallet.cs.umass.edu/} topic modelling package has been utilized to group document into $ m_d $ topics. Number of documents that have been selected as corresponding documents to candidate expert $ c_i $, $|D^{c_i}|$, and query $ q_j $, $|D^{q_j}|$, has been set to $ 2000 $ (i.e. $ n = 2000 $) for Java and C\# domain. For Android domain, the effect of parameter $n$ has been investigated in the quality of results. Also, we have considered two kernels, $ k = 2 $, in the both convolution neural networks (i.e. $ \chi_c \text{ and } \chi_q $) of the proposed model. The size of candidate and query latent vectors, $ m_c $ and $ m_q $, has been set to $ 32 $. 

\subsection{Experimental Results and Analysis}
\label{experimental_results_and_analysis}

In this section, we have compared the proposed method against five well-known methods in expert finding literature. To investigate the quality of ranking, we have utilized three prominent evaluation metrics, including NDCG, MRR, and ERR. 

\noindent $ \blacktriangleright $ \textbf{Analysis of Parameter $ n $:} At first, we have conducted  \colorbox{\hltcolor}{a set of experiments} to investigate the effect of number of documents related to queries (i.e. $n$) on the effectiveness of the proposed method. Figures \ref{fig:lda_metrics}-(a) to (c) shows \colorbox{\hltcolor}{the sensitivity} analysis of the proposed method in Android domain when the parameter $n$ varies. As these figures show, the proposed method doesn't perform well for small values of $n$ because of the lack of expertise evidence in small amount of documents. Although the proposed method performs better for bigger values of $n$, the model performance for small values of $n$ is better than some baselines. Another noteworthy point is that by increasing the value of $n$ the quality of results increases unless for large values of $n$.  \colorbox{\hltcolor}{It can be explained by that} the high increase in the value of $n$ makes the inputs of CNN layers to be sparse and decreases the quality of results consequently.   

\def\sc{0.33}
\begin{figure}
	\centering
	\begin{tabular}{c}
	     	\includegraphics [scale=\sc] {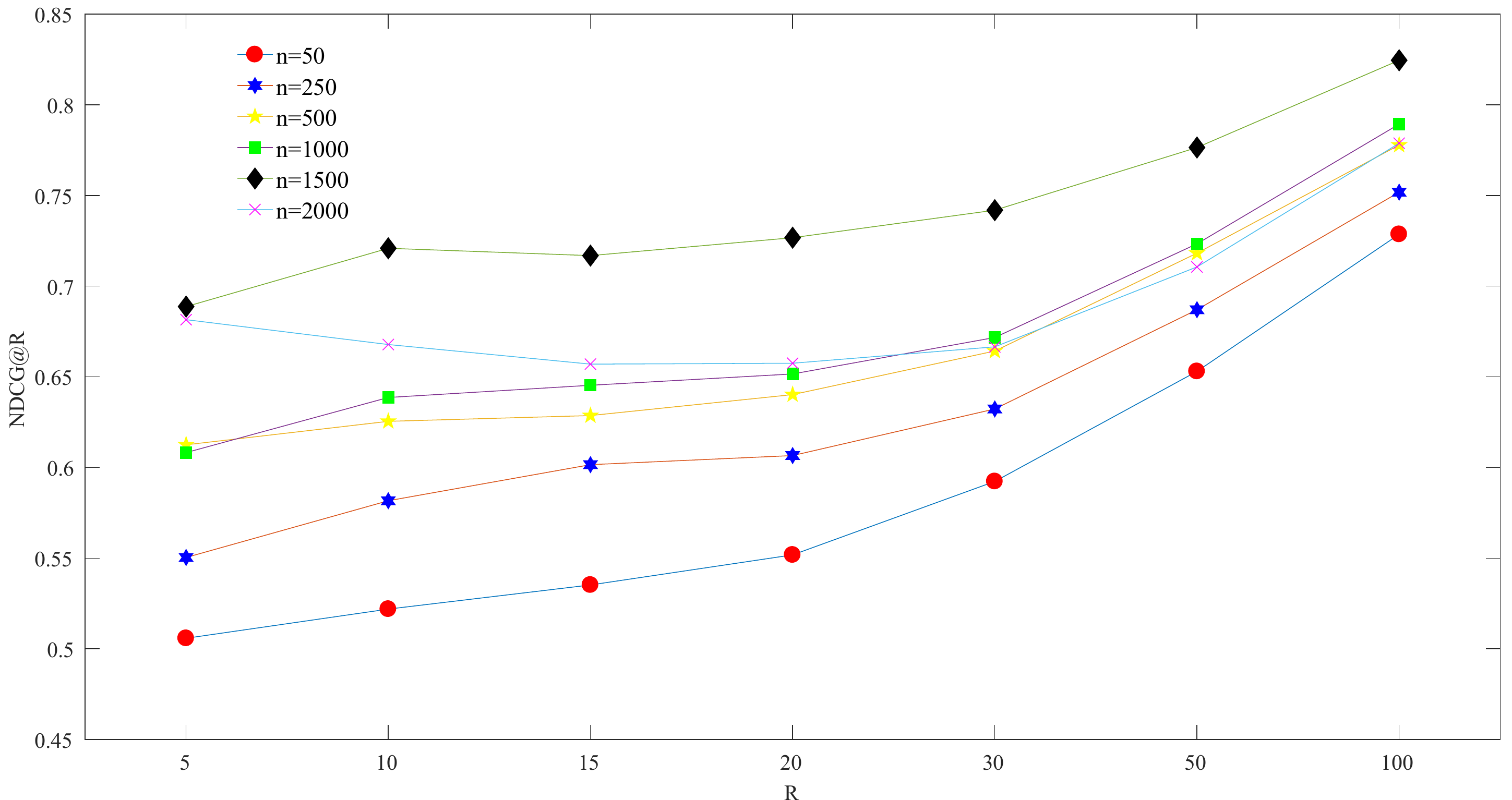}\\
	     	(a) NDCG \\ \\
            \includegraphics [scale=\sc] {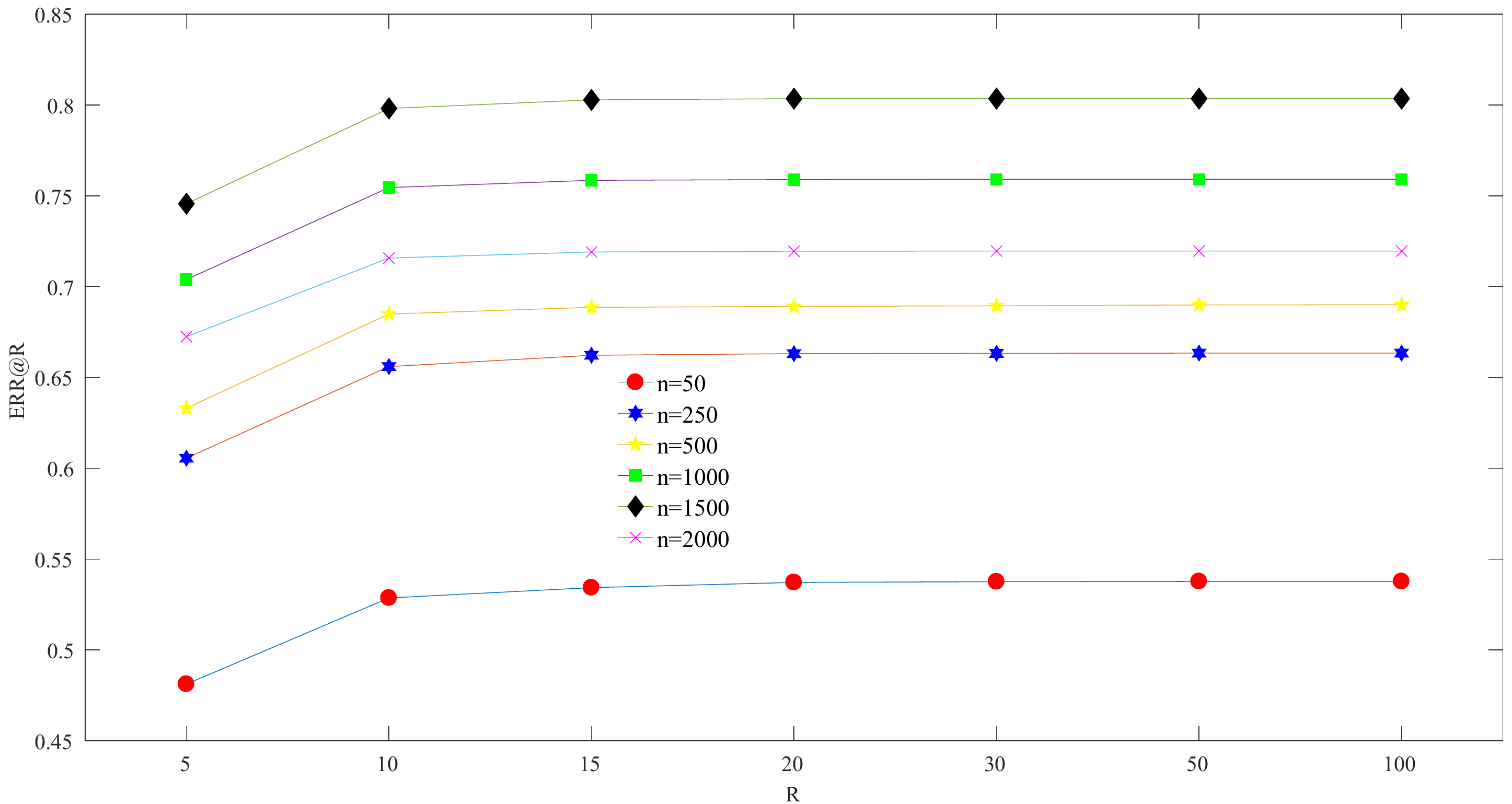}\\
            (b) ERR \\ \\
	        \includegraphics [scale=\sc] {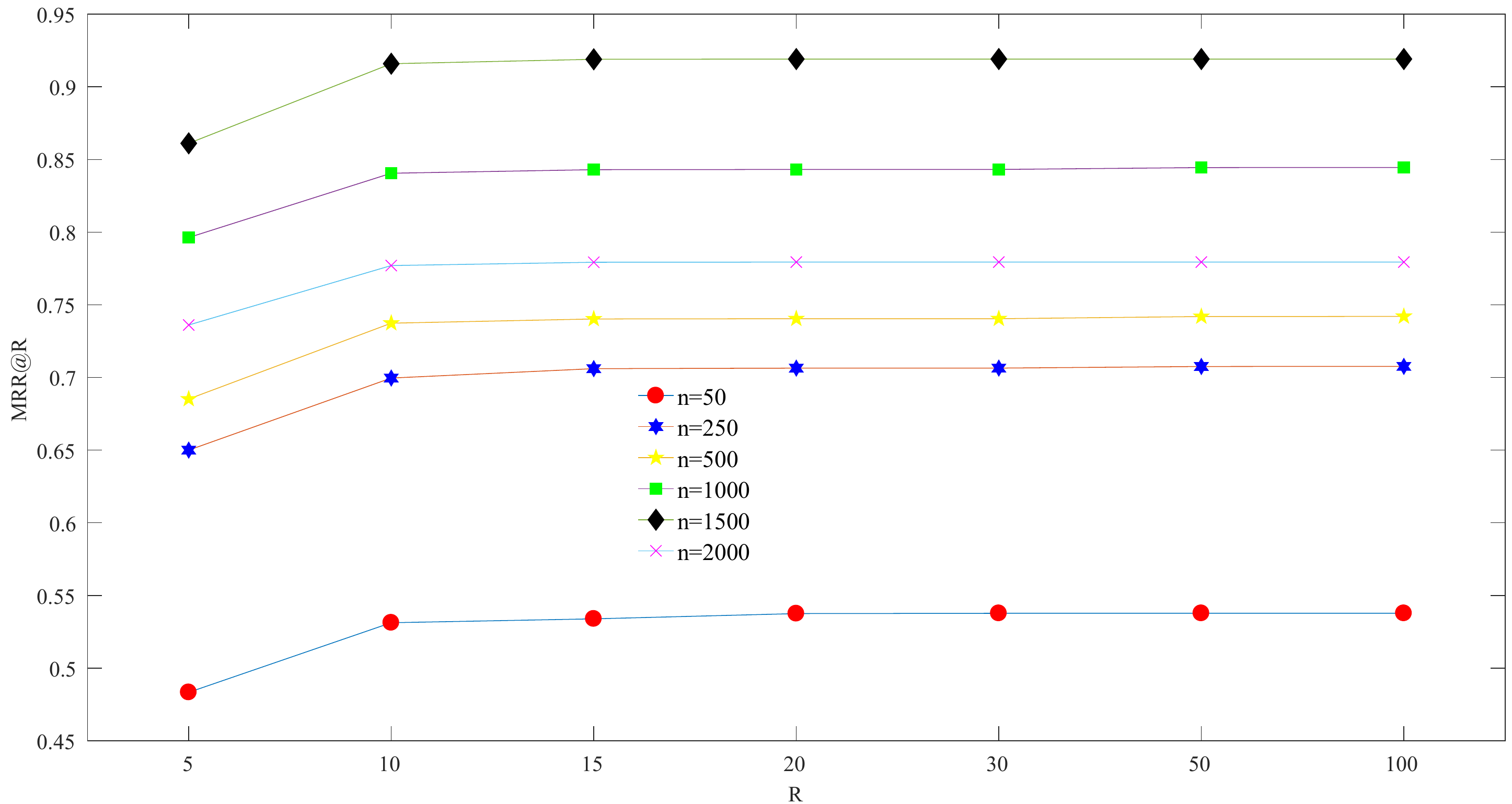}\\
	        (c) MRR 
	\end{tabular}
	\caption{Sensitivity analysis of the proposed method in Android domain when the parameter $n$ varies.}
	\label{fig:lda_metrics}
\end{figure}

\noindent $ \blacktriangleright $  \textbf{\colorbox{\hltcolor}{Analyzing} the Effect of Document Embedding:} In this section, we have investigated the effect of two different document embedding approaches proposed in \cite{le2014distributed,blei2003latent} on the quality of results. Tables \ref{tab:proposed_method_Java}, \ref{tab:proposed_method_Android} and \ref{tab:proposed_method_Csharp} compares the quality of ranking for two different document embedding approaches against baselines on three Java, Android, and C\# domains, respectively. As these tables show, our proposed method has greatly improved the quality of ranking in all three domains. The most important reason behind this improvement is extracting text features from documents by CNN to match users and queries semantically. Also, our proposed method detects more T-shaped experts in top of the ranked list than baselines for two reasons. First, it performs better than baselines in terms of MRR evaluation metric. Second, it improves baselines in terms of our evaluation metrics for the small values of $ R $. To better understand the effect of \colorbox{\hltcolor}{parameter} $ R $  in the quality of ranking, we have compared the proposed method against baselines in Figures \ref{fig:java_domain}, \ref{fig:android_domain}, and \ref{fig:csharp_domain}. 

\begin{table}
	\caption{Comparison of the proposed methods against baselines in Java domain.}
	\centering
	\begin{center}
		\begin{tabular}{llccccc|cc}
			\noalign{\hrule height 1pt}
			& &\multicolumn{5}{c}{\textbf{Baselines}} &\multicolumn{2}{c}{\textbf{Proposed}}\\
			\cline{3-9}
			$ \boldsymbol{R} $ & \textbf{Metric} &\rot{\textbf{DBA}\cite{balog2012expertise}}&\rot{\textbf{EBA}\cite{gharebagh2018t}}&\rot{\textbf{XEBA}\cite{gharebagh2018t}}&\rot{\textbf{LSTM-BFS}\cite{dehghan2019temporal}}&\rot{\textbf{LSTM-DFS}\cite{dehghan2019temporal}}&\rot{\textbf{Doc2vec}}& \rot{\textbf{LDA}}\\
			\noalign{\hrule height 1pt}
			& NDCG@R &$ 37.0 $&$ 35.2 $&$ 42.0 $&$ 65.2 $&$ 68.4 $&$ 60.7 $&$ 61.7 $\\
            $ 5 $ & ERR@R &$ 36.5 $&$ 39.4 $&$ 45.1 $&$ 66.2 $&$ 68.3 $&$ 62.4 $&$ 66.1 $\\
            & MRR@R&$ 17.1 $&$ 29.5 $&$ 37.5 $&$ 74.0 $&$ 73.4 $&$ 66.7 $&$ 71.2 $\\
            \noalign{\hrule height 1pt}
            & NDCG@R &$ 35.6 $&$ 33.2 $&$ 39.7 $&$ 65.3 $&$ 68.4 $&$ 60.2 $&$ 63.7 $\\
            $ 10 $ & ERR@R &$ 39.3 $&$ 42.3 $&$ 47.3 $&$ 67.1 $&$ 67.2 $&$ 65.6 $&$ 69.9 $\\
            & MRR@R&$ 19.7 $&$ 32.2 $&$ 40.4 $&$ 74.0 $&$ 73.4 $&$ 69.2 $&$ 74.9 $\\
            \noalign{\hrule height 1pt}
            & NDCG@R &$ 34.3 $&$ 32.6 $&$ 38.0 $&$ 63.9 $&$ 67.3 $&$ 59.5 $&$ 65.0 $\\
            $ 15 $ & ERR@R &$ 39.8 $&$ 42.9 $&$ 47.8 $&$ 67.0 $&$ 67.2 $&$ 66.0 $&$ 70.1 $\\
            & MRR@R&$ 20.4 $&$ 33.5 $&$ 41.0 $&$ 74.0 $&$ 73.4 $&$ 69.6 $&$ 75.0 $\\
            \noalign{\hrule height 1pt}
            & NDCG@R &$ 32.7 $&$ 31.7 $&$ 36.2 $&$ 62.7 $&$ 67.1 $&$ 59.1 $&$ 66.4 $\\
            $ 20 $ & ERR@R &$ 40.0 $&$ 43.1 $&$ 47.9 $&$ 67.0 $&$ 67.3 $&$ 66.1 $&$ 70.1 $\\
            & MRR@R&$ 20.8 $&$ 33.7 $&$ 41.2 $&$ 74.0 $&$ 73.4 $&$ 69.6 $&$ 75.0 $\\
            \noalign{\hrule height 1pt}
            & NDCG@R &$ 31.1 $&$ 30.1 $&$ 35.7 $&$ 61.5 $&$ 65.0 $&$ 60.5 $&$ 69.0 $\\
            $ 30 $ & ERR@R &$ 40.1 $&$ 43.2 $&$ 48.0 $&$ 67.2 $&$ 67.8 $&$ 66.1 $&$ 70.1 $\\
            & MRR@R&$ 21.0 $&$ 33.7 $&$ 41.4 $&$ 74.0 $&$ 73.4 $&$ 69.6 $&$ 75.0 $\\
            \noalign{\hrule height 1pt}
            & NDCG@R &$ 30.5 $&$ 29.1 $&$ 34.9 $&$ 59.7 $&$ 63.2 $&$ 66.2 $&$ 73.6 $\\
            $ 50 $ & ERR@R &$ 40.2 $&$ 43.2 $&$ 48.0 $&$ 65.8 $&$ 67.8 $&$ 66.1 $&$ 70.1 $\\
            & MRR@R&$ 21.3 $&$ 33.8 $&$ 41.5 $&$ 74.0 $&$ 73.4 $&$ 69.6 $&$ 75.0 $\\
            \noalign{\hrule height 1pt}
            & NDCG@R &$ 31.4 $&$ 29.3 $&$ 35.8 $&$ 59.6 $&$ 62.8 $&$ 74.2 $& \underline{$ \boldsymbol{80.8} $}\\
            $ 100 $ & ERR@R &$ 40.2 $&$ 43.2 $&$ 48.0 $&$ 65.6 $&$ 68.4 $&$ 66.1 $& \underline{$ \boldsymbol{70.1} $}\\
            & MRR@R&$ 21.5 $&$ 33.9 $&$ 41.6 $&$ 74.0 $&$ 73.4 $&$ 69.6 $& \underline{$ \boldsymbol{75.0} $}\\
			\noalign{\hrule height 1pt}
		\end{tabular}
	\end{center}
	\label{tab:proposed_method_Java}
\end{table}

\begin{table}
	\caption{Comparison of the proposed methods against baselines in Android domain.}
	\centering
	\begin{center}
		\begin{tabular}{llccccc|cc}
			\noalign{\hrule height 1pt}
			& &\multicolumn{5}{c}{\textbf{BaseLines}} &\multicolumn{2}{c}{\textbf{Proposed}}\\
			\cline{3-9}
			$ \boldsymbol{R} $ & \textbf{Metric} &\rot{\textbf{DBA}\cite{balog2012expertise}}&\rot{\textbf{EBA}\cite{gharebagh2018t}}&\rot{\textbf{XEBA}\cite{gharebagh2018t}}&\rot{\textbf{LSTM-BFS}\cite{dehghan2019temporal}}&\rot{\textbf{LSTM-DFS}\cite{dehghan2019temporal}}&\rot{\textbf{Doc2vec}}& \rot{\textbf{LDA}}\\
			\noalign{\hrule height 1pt}
            & NDCG@R &$ 12.2 $&$ 23.5 $&$ 27.8 $&$ 48.8 $&$ 55.9 $&$ 50.8 $&$ 68.9 $\\
            $ 5 $ & ERR@R &$ 14.4 $&$ 27.9 $&$ 31.3 $&$ 54.3 $&$ 61.1 $&$ 53.9 $&$ 74.6 $\\
            & MRR@R&$ 9.6 $&$ 24.7 $&$ 26.9 $&$ 63.3 $&$ 68.5 $&$ 52.2 $&$ 86.1 $\\
            \noalign{\hrule height 1pt}
            & NDCG@R &$ 13.3 $&$ 23.3 $&$ 26.2 $&$ 49.1 $&$ 56.1 $&$ 55.5 $&$ 72.1 $\\
            $ 10 $ & ERR@R &$ 16.9 $&$ 30.5 $&$ 33.1 $&$ 55.8 $&$ 62.4 $&$ 57.7 $&$ 79.8 $\\
            & MRR@R&$ 11.0 $&$ 26.1 $&$ 27.6 $&$ 64.1 $&$ 69.4 $&$ 55.1 $&$ 91.6 $\\
            \noalign{\hrule height 1pt}
            & NDCG@R &$ 14.5 $&$ 22.3 $&$ 26.2 $&$ 49.1 $&$ 55.3 $&$ 57.3 $&$ 71.7 $\\
            $ 15 $ & ERR@R &$ 18.1 $&$ 31.2 $&$ 34.4 $&$ 55.9 $&$ 62.5 $&$ 58.1 $&$ 80.3 $\\
            & MRR@R&$ 11.5 $&$ 26.5 $&$ 29.6 $&$ 64.1 $&$ 69.4 $&$ 55.3 $&$ 91.9 $\\
            \noalign{\hrule height 1pt}
            & NDCG@R &$ 14.8 $&$ 21.8 $&$ 27.3 $&$ 49.2 $&$ 55.0 $&$ 58.8 $&$ 72.7 $\\
            $ 20 $ & ERR@R &$ 18.4 $&$ 31.4 $&$ 35.0 $&$ 55.9 $&$ 62.5 $&$ 58.1 $&$ 80.3 $\\
            & MRR@R&$ 11.9 $&$ 26.8 $&$ 30.5 $&$ 64.1 $&$ 69.4 $&$ 55.3 $&$ 91.9 $\\
            \noalign{\hrule height 1pt}
            & NDCG@R &$ 14.7 $&$ 21.0 $&$ 26.4 $&$ 49.4 $&$ 54.4 $&$ 60.7 $&$ 74.2 $\\
            $ 30 $ & ERR@R &$ 18.8 $&$ 31.9 $&$ 35.1 $&$ 55.9 $&$ 62.5 $&$ 58.2 $&$ 80.4 $\\
            & MRR@R&$ 12.1 $&$ 27.4 $&$ 30.5 $&$ 64.1 $&$ 69.4 $&$ 55.5 $&$ 91.9 $\\
            \noalign{\hrule height 1pt}
            & NDCG@R &$ 15.4 $&$ 20.8 $&$ 26.0 $&$ 51.3 $&$ 54.0 $&$ 65.6 $&$ 77.6 $\\
            $ 50 $ & ERR@R &$ 19.0 $&$ 32.0 $&$ 35.2 $&$ 55.8 $&$ 62.6 $&$ 58.2 $&$ 80.4 $\\
            & MRR@R&$ 12.6 $&$ 27.6 $&$ 30.7 $&$ 64.1 $&$ 69.4 $&$ 55.6 $&$ 91.9 $\\
            \noalign{\hrule height 1pt}
            & NDCG@R &$ 17.6 $&$ 23.2 $&$ 30.4 $&$ 52.7 $&$ 55.7 $&$ 73.5 $& \underline{$ \boldsymbol{82.5} $}\\
            $ 100 $ & ERR@R &$ 19.1 $&$ 32.0 $&$ 35.2 $&$ 55.4 $&$ 62.6 $&$ 58.2 $& \underline{$ \boldsymbol{80.4} $}\\
            & MRR@R&$ 12.9 $&$ 27.7 $&$ 30.7 $&$ 64.1 $&$ 69.4 $&$ 55.6 $& \underline{$ \boldsymbol{91.9} $}\\
			\noalign{\hrule height 1pt}
		\end{tabular}
	\end{center}
	\label{tab:proposed_method_Android}
\end{table}

\begin{table}
	\caption{Comparison of the proposed methods against baselines in C\# domain.}
	\centering
	\begin{center}
		\begin{tabular}{llccccc|cc}
			\noalign{\hrule height 1pt}
			& &\multicolumn{5}{c}{\textbf{BaseLines}} &\multicolumn{2}{c}{\textbf{Proposed}}\\
			\cline{3-9}
			$ \boldsymbol{R} $ & \textbf{Metric} &\rot{\textbf{DBA}\cite{balog2012expertise}}&\rot{\textbf{EBA}\cite{gharebagh2018t}}&\rot{\textbf{XEBA}\cite{gharebagh2018t}}&\rot{\textbf{LSTM-BFS}\cite{dehghan2019temporal}}&\rot{\textbf{LSTM-DFS}\cite{dehghan2019temporal}}&\rot{\textbf{Doc2vec}}& \rot{\textbf{LDA}}\\
			\noalign{\hrule height 1pt}
			& NDCG@R &$ 26.4 $&$ 30.7 $&$ 34.0 $&$ 61.4 $&$ 65.0 $&$ 66.4 $&$ 76.3 $\\
            $ 5 $ & ERR@R &$ 29.1 $&$ 34.6 $&$ 39.7 $&$ 67.9 $&$ 59.3 $&$ 65.8 $&$ 75.0 $\\
            & MRR@R&$ 17.7 $&$ 28.7 $&$ 34.3 $&$ 73.3 $&$ 67.7 $&$ 70.4 $&$ 82.1 $\\
            \noalign{\hrule height 1pt}
            & NDCG@R &$ 26.5 $&$ 28.6 $&$ 33.0 $&$ 61.6 $&$ 63.5 $&$ 66.3 $&$ 74.6 $\\
            $ 10 $ & ERR@R &$ 32.1 $&$ 36.9 $&$ 42.5 $&$ 68.7 $&$ 60.6 $&$ 69.7 $&$ 79.1 $\\
            & MRR@R&$ 20.5 $&$ 30.8 $&$ 37.1 $&$ 73.3 $&$ 67.7 $&$ 73.9 $&$ 86.2 $\\
            \noalign{\hrule height 1pt}
            & NDCG@R &$ 26.2 $&$ 28.7 $&$ 33.3 $&$ 59.2 $&$ 63.5 $&$ 63.4 $&$ 71.4 $\\
            $ 15 $ & ERR@R &$ 32.9 $&$ 37.6 $&$ 42.9 $&$ 69.0 $&$ 62.3 $&$ 69.9 $&$ 79.4 $\\
            & MRR@R&$ 22.1 $&$ 32.1 $&$ 38.2 $&$ 73.8 $&$ 67.7 $&$ 74.1 $&$ 86.4 $\\
            \noalign{\hrule height 1pt}
            & NDCG@R &$ 25.9 $&$ 28.5 $&$ 33.3 $&$ 59.4 $&$ 61.2 $&$ 62.6 $&$ 70.7 $\\
            $ 20 $ & ERR@R &$ 33.1 $&$ 37.9 $&$ 43.0 $&$ 68.8 $&$ 62.5 $&$ 69.9 $&$ 79.4 $\\
            & MRR@R&$ 22.8 $&$ 32.4 $&$ 38.2 $&$ 73.8 $&$ 67.9 $&$ 74.1 $&$ 86.4 $\\
            \noalign{\hrule height 1pt}
            & NDCG@R &$ 25.9 $&$ 28.2 $&$ 32.6 $&$ 58.7 $&$ 59.4 $&$ 63.4 $&$ 71.1 $\\
            $ 30 $ & ERR@R &$ 33.3 $&$ 38.1 $&$ 43.1 $&$ 68.4 $&$ 62.6 $&$ 70.0 $&$ 79.4 $\\
            & MRR@R&$ 22.8 $&$ 32.8 $&$ 38.2 $&$ 74.0 $&$ 67.9 $&$ 74.1 $&$ 86.4 $\\
            \noalign{\hrule height 1pt}
            & NDCG@R &$ 24.4 $&$ 26.5 $&$ 30.9 $&$ 57.6 $&$ 58.2 $&$ 66.3 $&$ 73.5 $\\
            $ 50 $ & ERR@R &$ 33.4 $&$ 38.1 $&$ 43.1 $&$ 68.4 $&$ 62.4 $&$ 70.0 $&$ 79.4 $\\
            & MRR@R&$ 23.0 $&$ 32.8 $&$ 38.2 $&$ 74.0 $&$ 68.1 $&$ 74.1 $&$ 86.4 $\\
            \noalign{\hrule height 1pt}
            & NDCG@R &$ 23.7 $&$ 26.2 $&$ 30.1 $&$ 57.9 $&$ 58.7 $&$ 76.1 $& \underline{$ \boldsymbol{82.2} $}\\
            $ 100 $ & ERR@R &$ 33.4 $&$ 38.1 $&$ 43.1 $&$ 68.4 $&$ 59.6 $&$ 70.0 $& \underline{$ \boldsymbol{79.4} $}\\
            & MRR@R&$ 23.0 $&$ 32.8 $&$ 38.2 $&$ 74.1 $&$ 68.3 $&$ 74.1 $& \underline{$ \boldsymbol{86.4} $}\\
			\noalign{\hrule height 1pt}
		\end{tabular}
	\end{center}
	\label{tab:proposed_method_Csharp}
\end{table}

\def\sc{0.33}
\begin{figure}
	\centering
	\begin{tabular}{c}
	     	\includegraphics [scale=\sc] {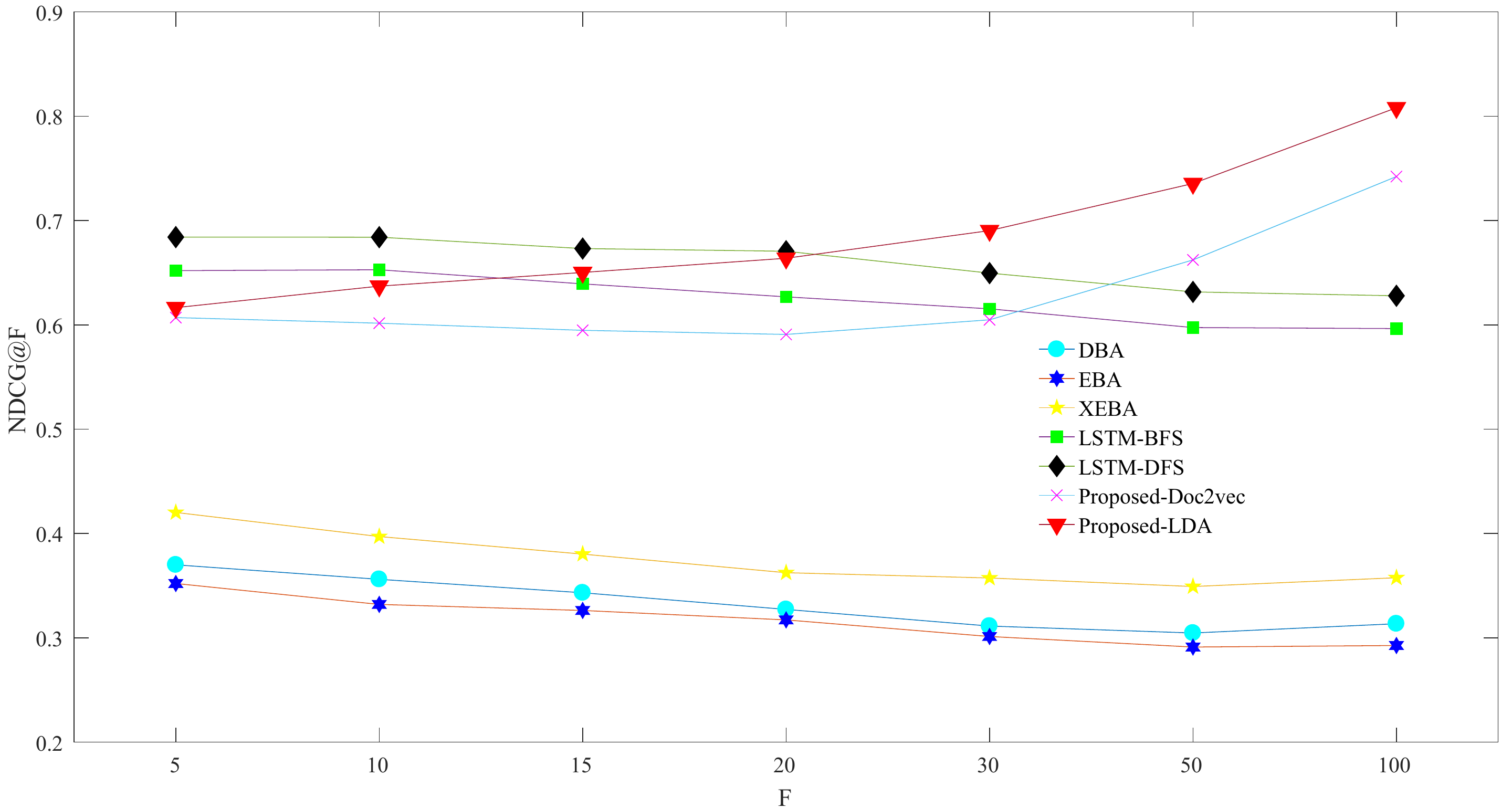}\\
	     	(a) NDCG \\ \\
            \includegraphics [scale=\sc] {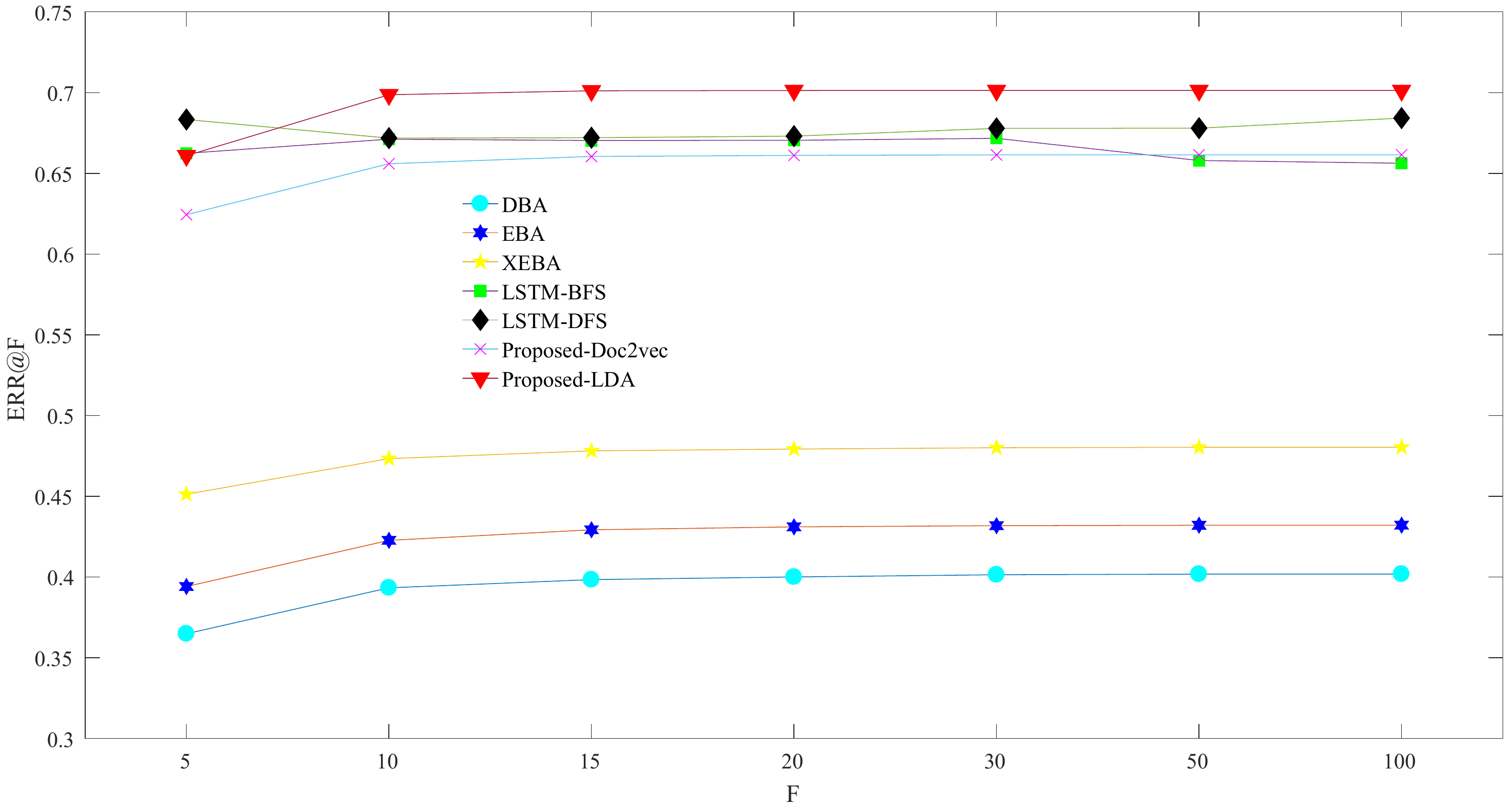}\\
            (b) ERR \\ \\
	        \includegraphics [scale=\sc] {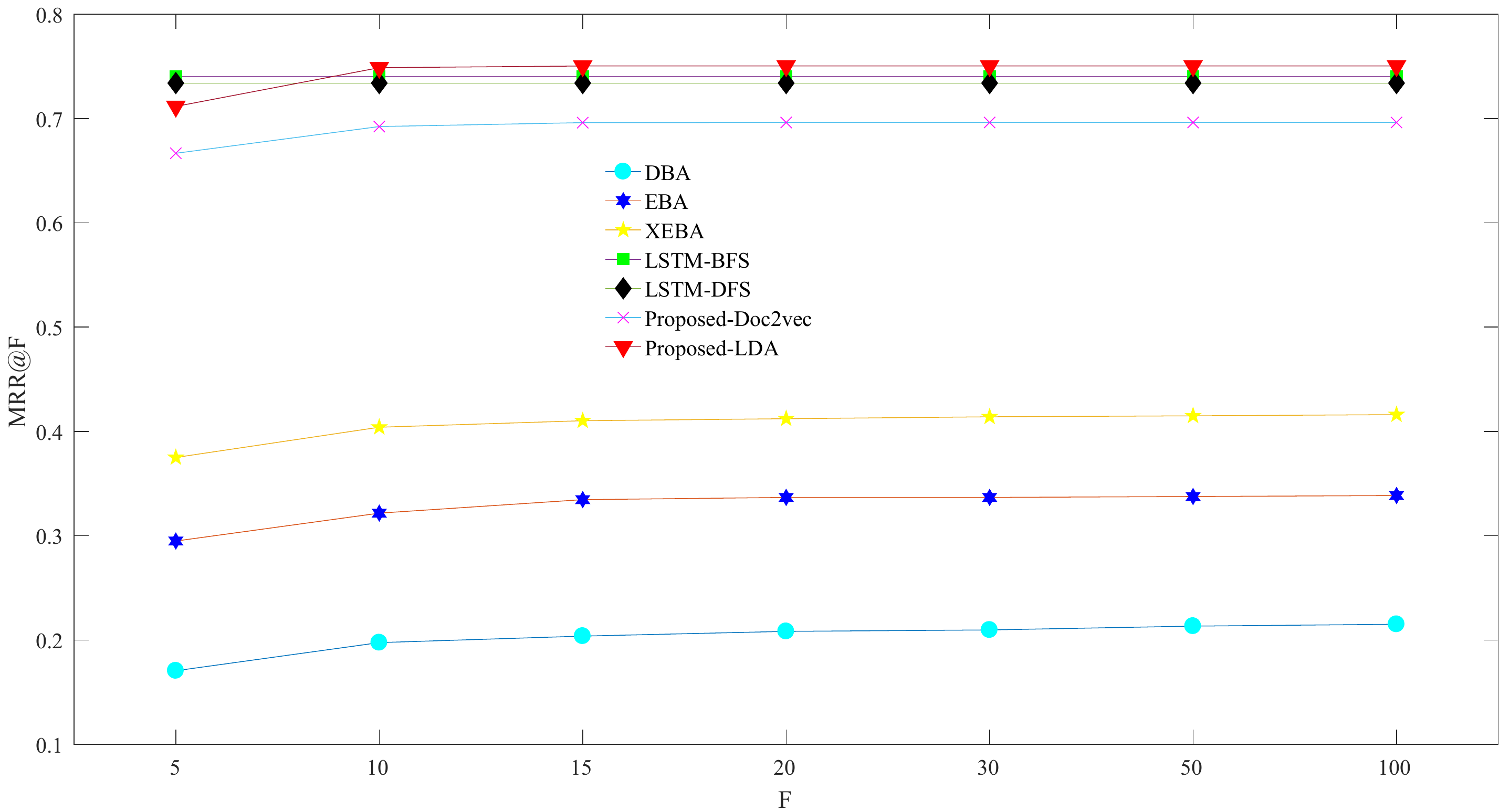}\\
	        (c) MRR 
	\end{tabular}
	\caption{Sensitivity analysis of the proposed method in Java domain when the parameter $R$ varies.}
	\label{fig:java_domain}
\end{figure}

\def\sc{0.33}
\begin{figure}
	\centering
	\begin{tabular}{c}
	     	\includegraphics [scale=\sc] {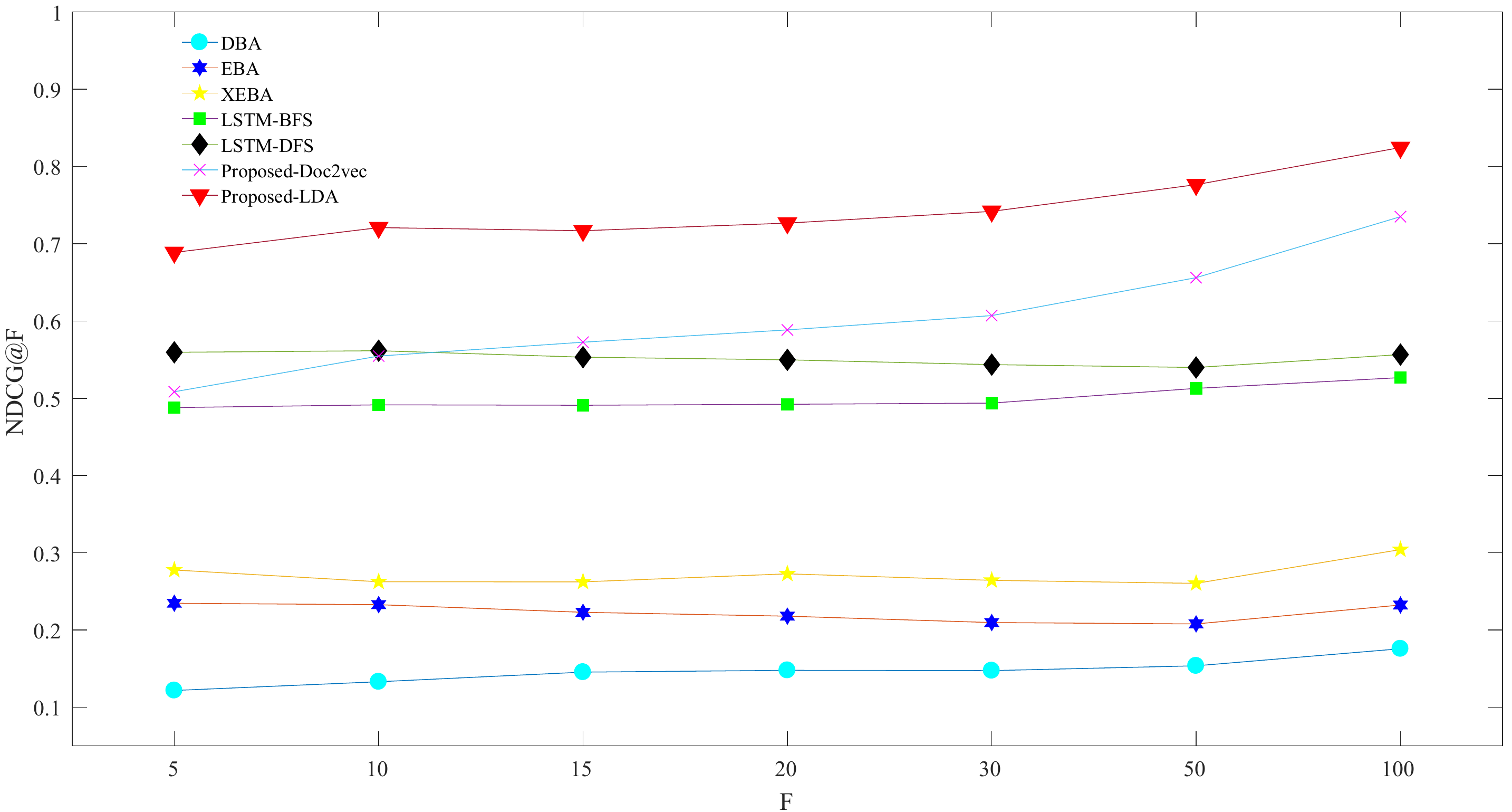}\\
	     	(a) NDCG \\ \\
            \includegraphics [scale=\sc] {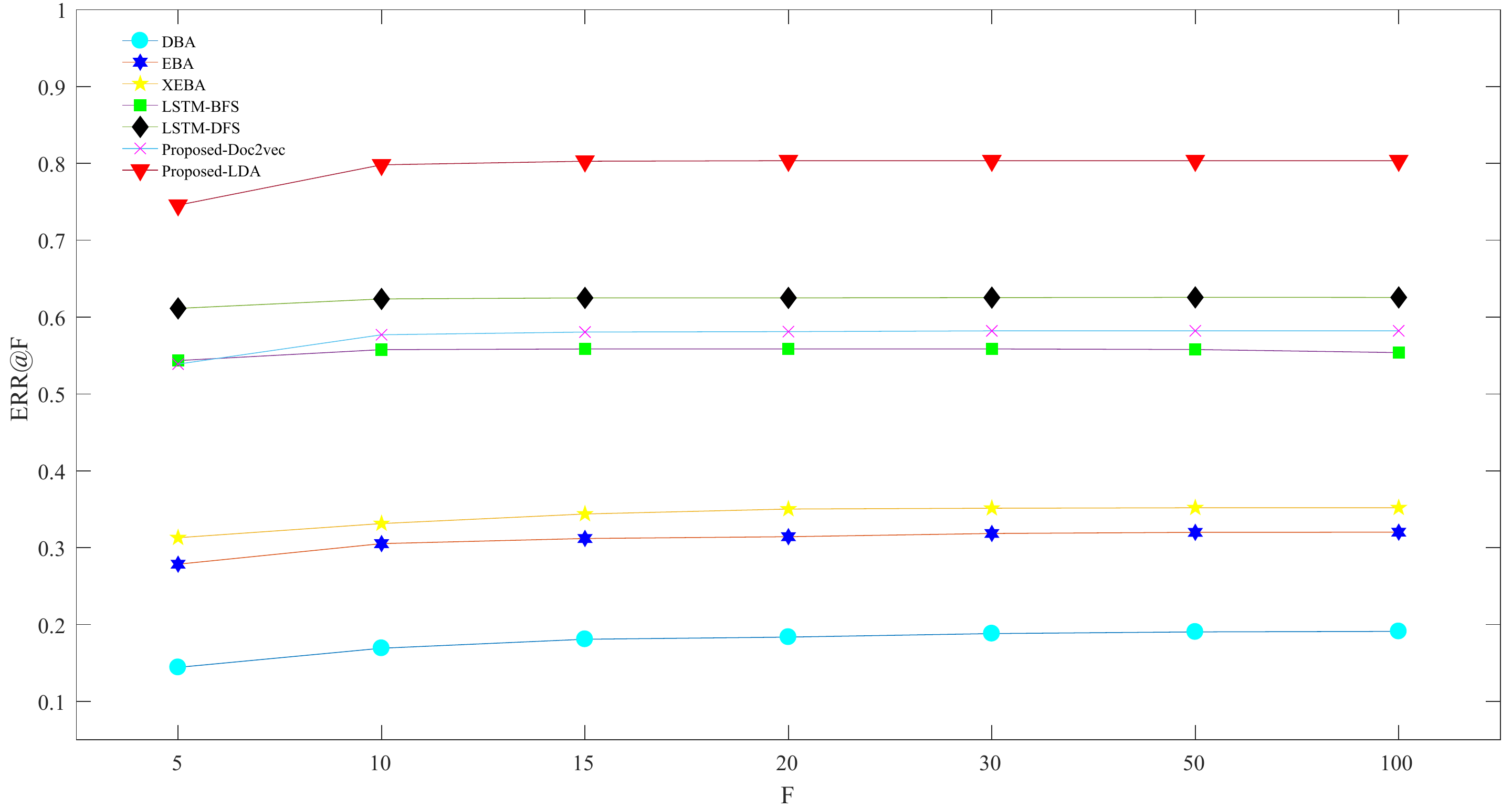}\\
            (b) ERR \\ \\
	        \includegraphics [scale=\sc] {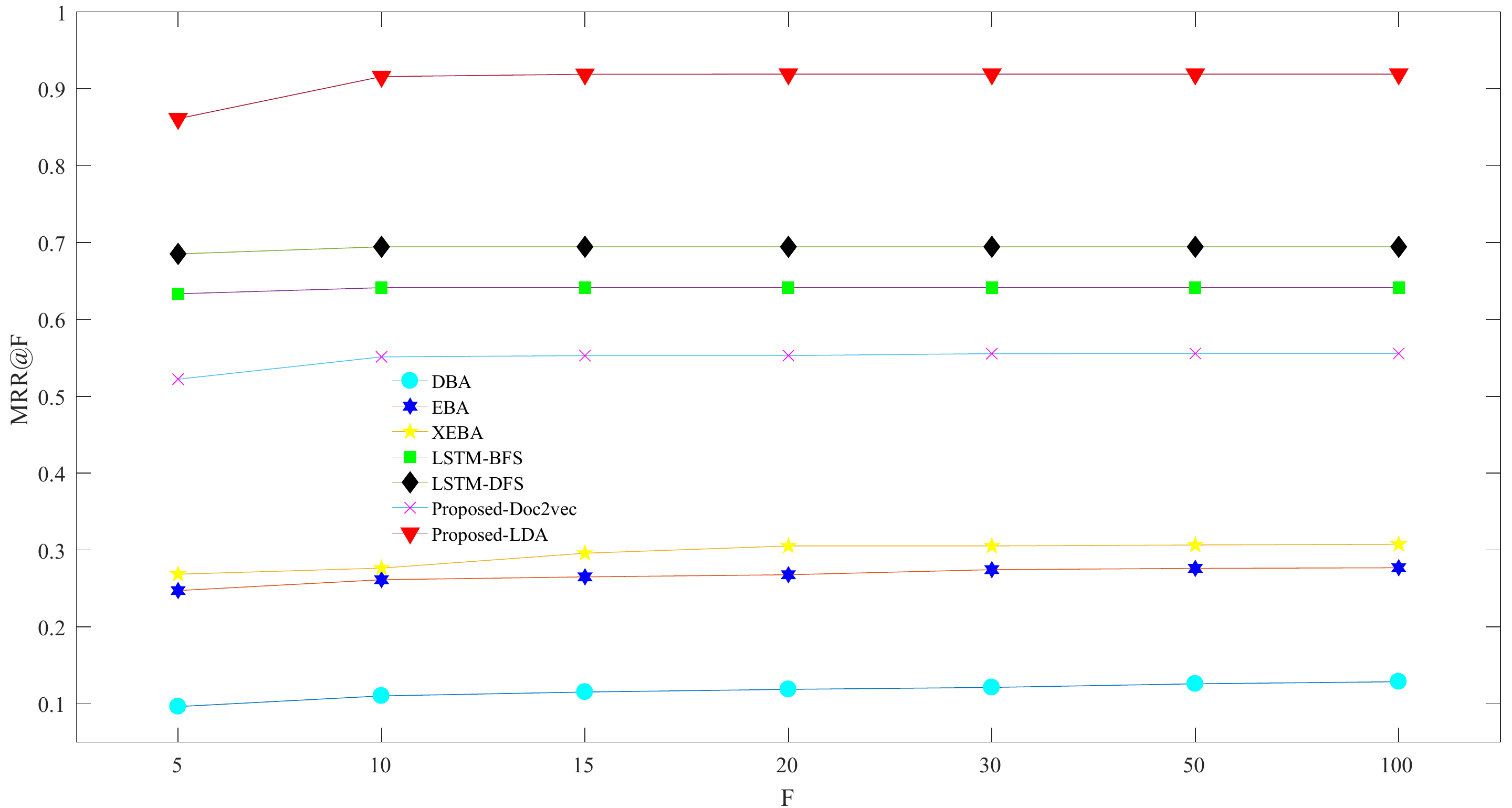}\\
	        (c) MRR 
	\end{tabular}
	\caption{Sensitivity analysis of the proposed method in Android domain when the parameter $R$ varies.}
	\label{fig:android_domain}
\end{figure}

\def\sc{0.33}
\begin{figure}
	\centering
	\begin{tabular}{c}
	     	\includegraphics [scale=\sc] {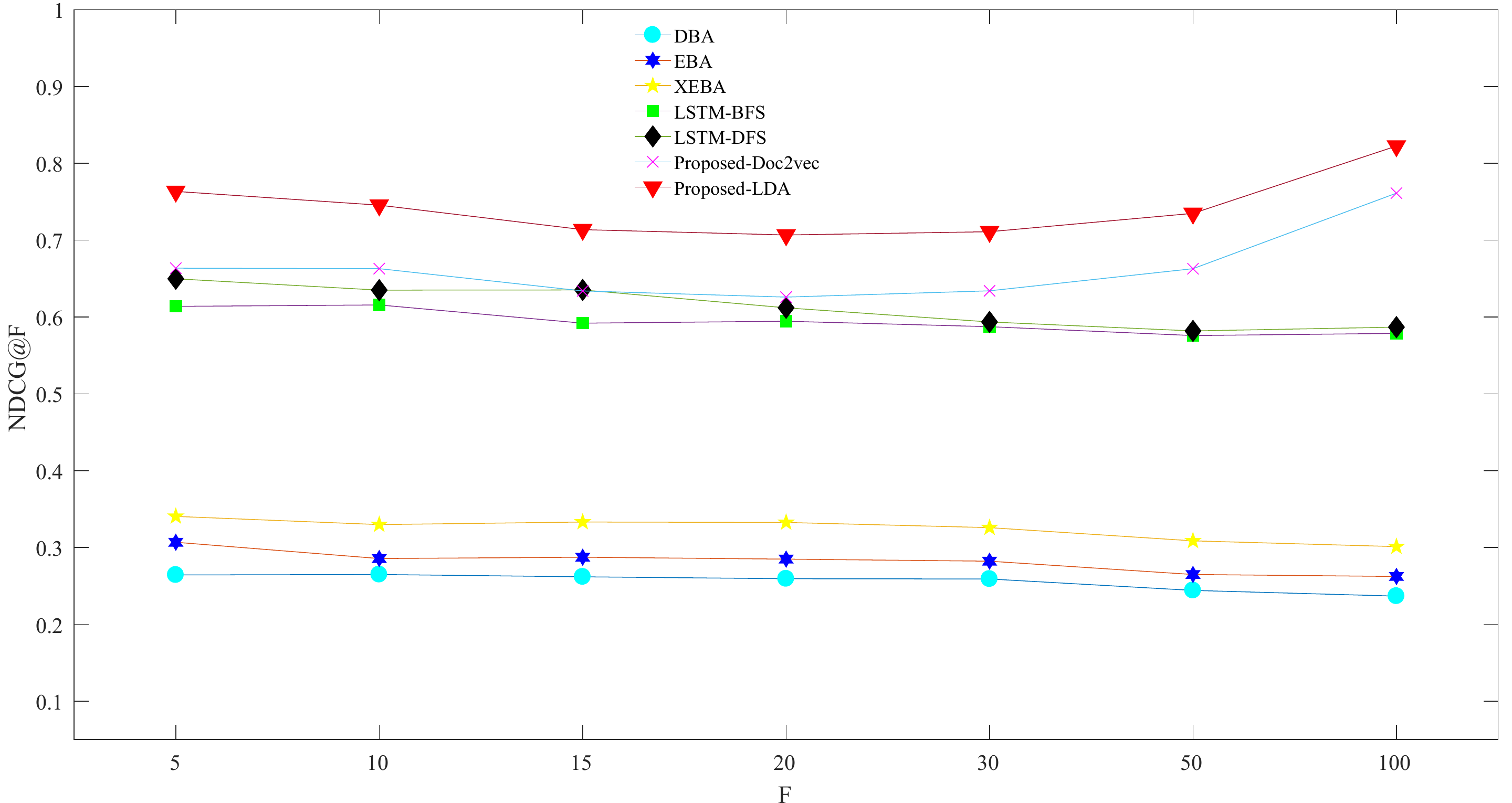}\\
	     	(a) NDCG \\ \\
            \includegraphics [scale=\sc] {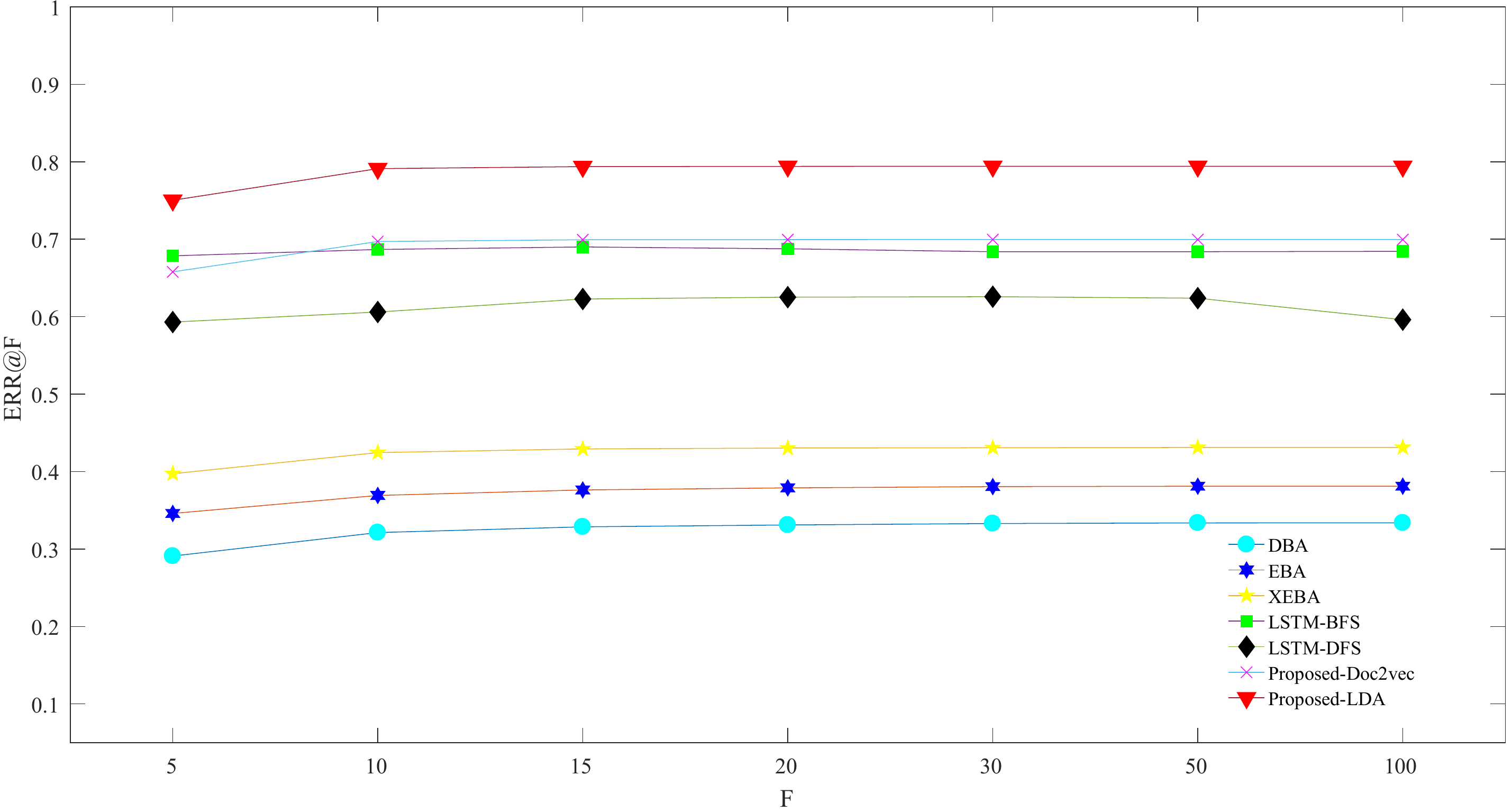}\\
            (b) ERR \\ \\
	        \includegraphics [scale=\sc] {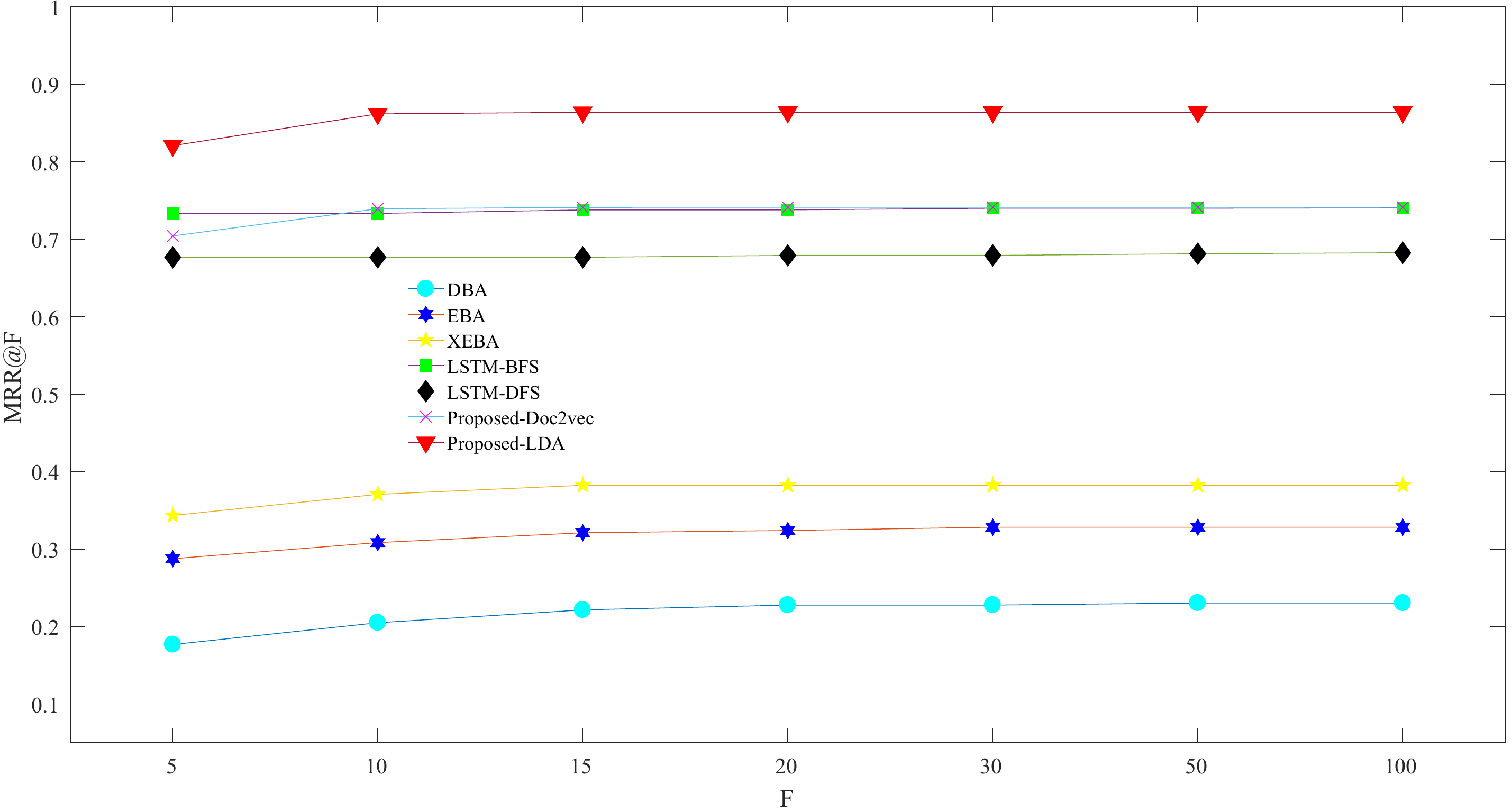}\\
	        (c) MRR 
	\end{tabular}
	\caption{Sensitivity analysis of the proposed method in C\# domain when the parameter $R$ varies.}
	\label{fig:csharp_domain}
\end{figure}

\noindent $ \blacktriangleright $ \textbf{Statistical comparison:} Finally, we have conducted a paired $t-$test with significance level $ 0.05 $. We have provided statistical comparison of the proposed method against baselines by using a paired $t-$test in Table \ref{table:paired_t_test}.

\begin{table}
	\centering
	\caption{Statistical comparison of the proposed method against baselines using paired $ t-$test. $x\sim y$ stands for that two methods $x$ and $y$ does not differ significantly.}
   		\begin{tabular}{lll}
		\noalign{\hrule height 1pt}
		\textbf{Domain} & \textbf{Metric} & \textbf{Paired $ t- $test} \\
		\noalign{\hrule height 1pt}
		& NDCG@100 & EBA$ < $DBA$< $XEBA$<$LSTM-BFS$<$LSTM-DFS$<$P-Doc2vec$<$P-LDA\\ 
		Java & ERR@100 & DBA$<$EBA$<$XEBA$<$LSTM-BFS$<$P-Doc2vec$<$LSTM-DFS$<$P-LDA\\ 
		& MRR@100 & DBA$<$EBA$<$XEBA$<$P-Doc2vec$<$LSTM-DFS$<$LSTM-BFS$<$P-LDA\\ 
		\cmidrule{1-3}
		& NDCG@100 & DBA$<$EBA$<$XEBA$<$LSTM-BFS$<$LSTM-DFS$<$P-Doc2vec$<$P-LDA\\ 
		Android & ERR@100 & DBA$<$EBA$<$XEBA$<$LSTM-BFS$<$P-Doc2vec$<$LSTM-DFS$<$P-LDA\\ 
		& MRR@100 & DBA$<$EBA$<$XEBA$<$P-Doc2vec$<$LSTM-BFS$<$LSTM-DFS$<$P-LDA\\  
		\cmidrule{1-3}
		& NDCG@100 & DBA$< $EBA$< $XEBA$< $LSTM-BFS$< $LSTM-DFS$< $P-Doc2vec$< $P-LDA\\ 
		C\# & ERR@100 & DBA$< $EBA$< $XEBA$< $LSTM-DFS$< $LSTM-BFS$< $P-Doc2vec$< $P-LDA\\ 
		& MRR@100 & DBA$< $EBA$< $XEBA$< $LSTM-DFS$< $LSTM-BFS$\sim $P-Doc2vec$< $P-LDA\\  
    	\cmidrule{1-3}
			\multicolumn{3}{|l|}{\textbf{P} in \textbf{P}-Doc2vec and \textbf{P}-LDA stands for the proposed method.} \\
		\noalign{\hrule height 1pt}
	\end{tabular}

	\label{table:paired_t_test}
	\end{table}

\section{Conclusion} \label{conclusion}
Community Question Answering networks provide valuable platforms for user to exchange knowledge or information. Stack Overflow network is an outstanding CQA providing \colorbox{\hltcolor}{a huge} source of knowledge that can be utilized by recruiters to hire new staff. Hiring new staff with reasonable cost is a vital concern of many companies. Forming a team consisting of T-shaped experts for software development is an excellent solution that optimizes company costs. This work proposed a new model for T-shaped expert finding that is based on convolutional neural networks. We proposed a model for T-shaped expert finding that learns latent vectors of users and queries from their corresponding documents simultaneously. Then, these latent vectors are utilized to match users and queries semantically. T-shaped experts grouping is a future direction to the proposed method.

\section*{Acknowledgement}
This research is funded by Vietnam National University, Hanoi (VNU) under project number QG.18.40.

\section*{References}
\bibliography{bibliography}

\end{document}